%% file: Journal_MIMO_final.tex
\documentclass[journal,onecolumn,final]{IEEEtran}
\usepackage{cite,acronym,paralist,xspace,stfloats}
\usepackage{amsmath,amssymb,amsfonts,amsthm}
\usepackage{algorithmic}
\usepackage{blkarray}
\usepackage{graphicx}
\usepackage{textcomp}
\usepackage{xcolor}
\usepackage{stmaryrd} 
\usepackage{dsfont}
\usepackage{etoolbox}

\input{arcom_functions.tex}
\input{arcom_notation.tex}
\input{arcom_acronyms.tex}

\usepackage[hidelinks]{hyperref}
\usepackage[nameinlink]{cleveref}

\newcounter{numrellocal}
\renewcommand{\thenumrellocal}{\alph{numrellocal}}
\newcounter{numrelglobal}

\makeatletter
\newcommand{\labrel}[2]{
  \stepcounter{numrellocal}
  \refstepcounter{numrelglobal}
  \ltx@label[equation]{#2}
  \stackrel{\textnormal{(\thenumrellocal)}}{\mathstrut{#1}}
}
\makeatother
\everydisplay\expandafter{\the\everydisplay\setcounter{numrellocal}{0}} 
\newtheorem{theorem}{Theorem}%
\newtheorem{proposition}[theorem]{Proposition}%
\newtheorem{lemma}[theorem]{Lemma}%
\newtheorem{definition}[theorem]{Definition}%
\newtheorem{remark}{Remark}
\Crefname{lemma}{Lemma}{Lemmas}
\crefname{lemma}{lemma}{lemmas}

\def\BibTeX{{\rm B\kern-.05em{\sc i\kern-.025em b}\kern-.08em
    T\kern-.1667em\lower.7ex\hbox{E}\kern-.125emX}}

\ifCLASSINFOpdf
\else
\fi

\begin{document}
%
\title{Covert MIMO Communications \\under Variational Distance Constraint
\thanks{This work was supported by the National Science Foundation awards 1527387 and 1910859.}}
%
%
%

\author{Shi-Yuan Wang,~\IEEEmembership{Student Member,~IEEE}, and
        Matthieu R. Bloch,~\IEEEmembership{Senior Member,~IEEE}
\thanks{The authors are with the School
of Electrical and Computer Engineering, Georgia Institute of Technology, Atlanta,
GA, 30332 USA (e-mail: shi-yuan.wang@gatech.edu; matthieu.bloch@ece.gatech.edu).}
}

\maketitle

\begin{abstract}
The problem of covert communication over \acf{MIMO} \acf{AWGN} channels is investigated, in which a transmitter attempts to reliably communicate with a legitimate receiver while avoiding detection by a passive adversary. The covert capacity of the \ac{MIMO} \ac{AWGN} channel is characterized under a variational distance covertness constraint when the \ac{MIMO} channel matrices are static and known. The characterization of the covert capacity is also extended to a class of channels in which the legitimate channel matrix is known but the adversary's channel matrix is only known up to a rank and a spectral norm constraint.
\end{abstract}


%
\IEEEpeerreviewmaketitle

\section{Introduction}
%
%
%
%
\label{sec:intro}

Covert communications, also known as communications with low probability of detection, have long been used to transmit sensitive information without raising suspicion. While technologies such as spread-spectrum communications have been widely deployed, the information-theoretic limits of covert communication had not been investigated until recently. Much of the interest has been spurred by the discovery of a \emph{square-root law}~\cite{Bash2013a}, which limits the scaling with the coding blocklength $n$ of the number of reliable and covert communication bits over memoryless channels to $\calO(\sqrt{n})$. In other words, the standard capacity of covert communications is zero but the number of bits still grows with the blocklength. The optimal constant behind the $\calO(\sqrt{n})$ scaling then plays the role of the \emph{covert capacity} and has been characterized for many channels, including \acp{DMC} and \ac{AWGN} channels, using both relative entropy~\cite{Wang2016a,Bloch2016} and variational distance~\cite{Tahmasbi2019,Zhang2019a} as a covertness metric. Covert communications often require secret keys as an enabling resource, the amount of which can be characterized~\cite{Bloch2016}; in particular, no secret keys are required when the legitimate receiver obtains better observations than the adversary~\cite{Che2013}. Refined characterizations of the message and key sizes for finite length~\cite{Yan2019b,Yu_2021} and second-order asymptotics~\cite{Tahmasbi2019} are also known, although they are often not complete. Recent advances include the characterization of the covert capacity in network information theory problems~\cite{Arumugam2019,Arumugam2019a,Tan2019,Cho2021}, quantum channels~\cite{Bash2015,Wang2016b}, low-complexity code constructions~\cite{Freche2017,Kadampot2018,Kadampot2019,Lamarca2019,Zhang2020}, and system-level considerations highlighting how to allocate resources in the presence of covertness constraints~\cite{Hu2018a,Zheng2019}.  Particularly relevant to the present work, there have been attempts at studying \ac{MIMO}-\ac{AWGN} channels when measuring covertness using relative entropy as a covertness metric~\cite{Abdelaziz2017,Bendary2019,Bendary2020}.

Covertness must be measured in terms of a metric that captures how different the statistics of the observations are in presence and in absence of communication. Relative entropy has been a popular choice~\cite{Wang2016a,Bloch2016} because of its convenient analytical properties; 
however,  variational distance is the metric that is operationally relevant to the performance of the adversary's detector~\cite{Tahmasbi2019}. 
In this work, we therefore use variational distance to measure covertness, which requires specific techniques, especially in the converse proof.


The contributions of the present work are twofold.
\begin{inparaenum}[1)]
\item We revisit the \ac{MIMO}-\ac{AWGN} channel model of \cite{Abdelaziz2017,Bendary2019,Bendary2020} and, under the assumption that 
  \textcolor{black}{the null space of the main and adversary's channel matrices are trivial}, we obtain a closed-form 
  \textcolor{black}{of the covert capacity} with variational distance as the covertness metric. Our approach extends the techniques developed in~\cite{Tahmasbi2019,Zhang2019a} and the crux of contribution is the converse proof.
\item We investigate the problem of covert communication over compound \ac{MIMO}-\ac{AWGN} channels, {in particular,} the situation in which the adversary's channel matrix is only known up to a rank constraint and a spectral norm constraint~\cite{Schaefer2015,Abdelaziz2017,Bendary2019,Bendary2020}. Our approach differs from the analysis in~\cite{Schaefer2015,Abdelaziz2017,Bendary2019,Bendary2020} and borrows ideas from~\cite{He2014} to avoid implicit constraints on the adversary's operation when dealing with uncountable compound channels. 
\end{inparaenum}
A preliminary version of these results was presented in~\cite{Wang2020} but without complete proofs. The present work offers self-contained and detailed proofs.

\section{Channel Model}
\label{sec:not-sys}
\subsection{Notation}
Both $\log$ and $\exp$ should be understood in base $e$; hence, all information-theoretic quantities are nats. Calligraphic letters are used for sets and $|\cdot|$ denotes their cardinality. $\left(\cdot\right)^\dagger$ denotes the Moore-Penrose inverse of a matrix. \textcolor{black}{$\vct{M}\succeq\vct{0}$ denotes a positive semi-definite matrix $\vct{M}$.} $\avgH{\cdot}$, $\h{\cdot}$, $\avgI{\cdot;\cdot}$, and $\Hb{\cdot}$ denote the usual entropy, differential entropy, mutual information, and binary entropy function, respectively.

For a continuous alphabet $\Omega$ and any two distributions $P,~Q$ with densities $f_P,~f_Q$, respectively, the variational distance between $P$ and $Q$ is defined as
$\V{P, Q}\eqdef\frac{1}{2}\int_{\Omega}|f_P\left(x\right)-f_Q\left(x\right)|dx$ or equivalently 
$\V{P, Q}=\sup_{\calS\subseteq\Omega}|P\left(\calS\right)-Q\left(\calS\right)|$.
The relative entropy between $P$ and $Q$ is defined as 
$\avgD{P}{Q}\eqdef\int_{\Omega}f_P(x)\log\frac{f_P(x)}{f_Q(x)}dx$.
Pinsker's inequality ensures that $\V{P,Q}^2\leq\frac{1}{2}\min\left(\avgD{P}{Q},\avgD{Q}{P}\right)$. \textcolor{black}{Let $X\in\calX$ and $Y\in\calY$ be jointly distributed random variables acccording to $P\cdot W$, where $P$ has density $f_P$, and $W:(x, y)\mapsto W(y|x)$ is a transition probability from $\calX\to\calY$ with density $f_W$. We define the marginal distribution of $Y$ as $P\circ W$ with density $\int_{\calX}f_W(y|x)f_P(x)dx$.}

Moreover, for two integers $\floor{a}$ and $\ceil{b}$ such that $\floor{a}\leq \ceil{b}$, we define $\intseq{a}{b}\eqdef\{\floor{a}, \floor{a}+1, \cdots, \ceil{b}-1, \ceil{b}\}$; otherwise $\intseq{a}{b}\eqdef\emptyset$. For any $x\in\bbR$, we also define the $Q$-function $Q(x)\eqdef\int_x^\infty\frac{1}{\sqrt{2\pi}}e^{\frac{-x^2}{2}}dx$ and its inverse function $Q^{-1}(\cdot)$.

\subsection{System Model}
We consider a \ac{MIMO}-\ac{AWGN} channel in which a transmitter (Alice) with $N_a$ antennas attempts to reliably communicate with a legitimate receiver (Bob) with $N_b$ antennas in the presence of a passive adversary (the warden Willie) equipped with $N_w$ antennas. 
We assume that Bob and Willie possess more antennas than Alice, i.e., $N_a \leq N_b$ and $N_a \leq N_w$. Bob and Willie's received signals at every channel use are then
\begin{align}
\label{eq:MIMO-scheme-def}
    \vct{y} = \vct{H}_b\vct{x}+\vct{n}_b \quad\text{and}\quad
    \vct{z} = \vct{H}_w\vct{x}+\vct{n}_w,
\end{align}
respectively, where $\vct{x}\in\bbR^{N_a}$ is Alice's  transmitted signal and $\vct{H}_b$ and $\vct{H}_w$ are Bob's and Willie's channel matrices, assumed known to everyone. We further assume that both matrices have full rank, i.e., $m=\rk{\vct{H}_b}=\rk{\vct{H}_w}=N_a$. Hence, both channel matrices can be decomposed with a \ac{GSVD}~\cite{Khisti2010,Paige1981} as
\begin{equation}
    \begin{split}
        \vct{H}_b &= \vct{U}_b^\prime\vct{\Sigma}_b\vct{\Omega}^{-1}\vct{\Psi}^\hermittian = \vct{U}_b\vct{\Lambda}_b\vct{V}^\hermittian, \\
        \vct{H}_w &= \vct{U}_w^\prime\vct{\Sigma}_w\vct{\Omega}^{-1}\vct{\Psi}^\hermittian = \vct{U}_w\vct{\Lambda}_w\vct{V}^\hermittian,
    \end{split}
\end{equation}
where $\vct{\Psi}\in\bbR^{N_a\times N_a}$, $\vct{U}_b^\prime\in\bbR^{N_b\times N_b}$, and $\vct{U}_w^\prime\in\bbR^{N_w\times N_w}$ are orthogonal, $\vct{\Omega}\in\bbR^{m\times m}$ is lower triangular and nonsingular, and $\vct{V}^\hermittian\eqdef\vct{\Omega}^{-1}\vct{\Psi}^\hermittian$. Both $\vct{\Sigma}_b\in\bbR^{N_b\times m}$ and $ \vct{\Sigma}_w\in\bbR^{N_w\times m}$ are diagonal with positive elements, $\{\lambda_{b,j}\}_{j=1}^m$ and $\{\lambda_{w,j}\}_{j=1}^m$, respectively. We truncate $\vct{U}_b^\prime$ and $\vct{U}_w^\prime$ into $\vct{U}_b\in\bbR^{N_b\times m}$ and $\vct{U}_w\in\bbR^{N_w\times m}$, and define $\vct{\Lambda}_b=\diag{\{\lambda_{b,j}\}_{j=1}^m}$ and  $\vct{\Lambda}_w=\diag{\{\lambda_{w,j}\}_{j=1}^m}$. The noise vectors $\vct{n}_b\in\bbR$ and $\vct{n}_w\in\bbR$ are realizations of \ac{AWGN} distributed according to $\calN\left(\vct{0}, \sigma_b^2\vct{I}_{N_b}\right)$ and $\calN\left(\vct{0}, \sigma_w^2\vct{I}_{N_w}\right)$, respectively, assumed known to everyone. 

Furthermore, for $n\in\bbN^*$, we define the innocent symbol corresponding to the absence of communication as $\vct{x}_0=\vct{0}$; the output distributions induced by the innocent symbol at Bob and Willie are denoted $\Pinn\eqdef\calN\left(\vct{0}, \sigma_b^2\vct{I}_{N_b}\right)$ and $\Qinn\eqdef\calN\left(\vct{0}, \sigma_w^2\vct{I}_{N_w}\right)$, respectively. The associated product distributions are denoted by $\Pinnn=\prod_{i=1}^n\Pinn~\text{and}~\Qinnn=\prod_{i=1}^n\Qinn.$
\begin{remark}
  \label{rmk:full-rank}
We assume that both $\vct{H}_b$ and $\vct{H}_w$ have a trivial null space equal to $\{\vct{0}\}$. If this were not the case, the presence of a null space would result in the following scenarios. If $\vct{H}_w$ has a non-trivial null space, Alice can overcome the square-root law by steering her beam in the corresponding  directions~\cite{Abdelaziz2017,Bendary2019,Bendary2020}. If $\vct{H}_b$ has a non-trivial null space, Alice has no incentive to use the corresponding directions and would simply ignore them. We offer further discussion in Appendix~\ref{sec:furth-expl-full}.  
\end{remark}
\subsection{Problem Formulation}
Alice transmits a uniformly-distributed message $\rvW\in\intseq{1}{M_n}$ by encoding it into a codeword $\vct{X}^n=\matvct{\vct{X}}{1}{n}\in\bbR^{N_a\times n}$ of blocklength $n$ with the aid of a uniformly-distributed secret key $\rvS\in\intseq{1}{K_n}$ shared with Bob. The resulting code is called an $(n, M_n, K_n)$-code $\calC$, assumed known to everyone. Whether Alice communicates or not is controlled by $\phi\in\{0,1\}$, with $\phi=1$ indicating the transmission. Upon observing $\vct{Y}^n=\matvct{\vct{Y}}{1}{n}\in\bbR^{N_b\times n}$, Bob uses his knowledge of the secret key to form a reliable estimate $\widehat{\rvW}$ of $\rvW$. Reliability is measured by the maximal average probability of error
\begin{equation}
\label{eq:max-P-err}
    P_e^{(n)}\eqdef\max_{s}{\bar{P}^{(n)}_e(s)}+\P{\widehat{\phi}=1|\phi=0},
\end{equation}
where $\bar{P}^{(n)}_e(s)\eqdef\P{W\neq\widehat{W}|S=s,\phi=1}$, and we define $\bar{P}^{(n)}_e\eqdef\expect[S]{\P{W\neq\widehat{W}|S,\phi=1}}+\P{\widehat{\phi}=1|\phi=0}$.
In contrast, Willie's objective is to detect whether Alice is transmitting based on the observations $\vct{Z}^n=\matvct{\vct{Z}}{1}{n}\in\bbR^{N_w\times n}$ via a hypothesis test $T\left(\vct{Z}^n\right)$. In particular, Willie expects $\Qinnn$ when there is no transmission between Alice and Bob (i.e., the null hypothesis) and $\Qcoden$ when the transmission occurs (i.e., the alternative hypothesis), where $\Qcoden$ is the output distribution induced by the code $\calC$ used by Alice and Bob, $\forall ~\vct{z}^n\in\bbR^{N_w\times n},$
\begin{equation}
    \Qcoden\left(\vct{z}^n\right)=\frac{1}{M_nK_n}\sum_{\ell=1}^{M_n}\sum_{k=1}^{K_n}\Wwn\left(\vct{z}^n|\vct{x}^{(\ell k)n}\right).
\end{equation}
In the sequel, we use $\V{\smash{\widehat{Q}^n,\Qinnn}}$ as our covertness metric. When testing the null hypothesis $\Qinnn$ against the alternative hypothesis $\Qcoden$, any test $T\left(\vct{Z}^n\right)$ conducted by Willie on the observations $\vct{Z}^n$ satisfies 
$1\geq \alpha+\beta\geq1-\V{\Qcoden,\Qinnn},$ where $\alpha~\text{and}~\beta$ are the probabilities of false alarm and missed detection, respectively, and the lower bound can be achieved by an optimal test~\cite[Theorem 13.1.1]{lehmann2006testing}. In addition, the trade-off $\alpha+\beta=1$ is achieved with blind tests that do not use the observations. Consequently, making $\V{\Qcoden,\Qinnn}$ vanish amounts to rendering the adversary's hypothesis test effectively blind and hence achieves covertness. 

\begin{definition}
\label{def:cvt-tp-pair}
A 
reliable and covert throughput $r\in\bbR_+$ is achievable with corresponding key throughput $k\in\bbR_+$, if there exists a sequence of $(n, M_n, K_n, \delta)$-codes with increasing blocklength $n$ such that
\begin{align}
\label{eq:def-pair}
    \liminf{\frac{\log M_n}{\sqrt{n}d}}\geq r,\quad \limsup{\frac{\log M_nK_n}{\sqrt{n}d}}\leq r+k,
\end{align}
and 
\begin{align}
    \limn{P_e^{(n)}}=0, \quad\V{\Qcoden,\Qinnn}\leq\delta,
\end{align}
where $d=\invQfunc{\frac{1-\delta}{2}}$. The covert capacity $C_{\textnormal{covert}}$ is the supremum of achievable throughputs $r$.
\end{definition}
Note that, in our definition, we normalize the message and key size by $\sqrt{n}d$ instead of the usual choice, $n$; this is essential to unveil the square-root law behind the covertness and is justified a posteriori by the results in Section~\ref{sec:main-results}. Intuitively, the square-root law exists, for we are hiding messages in ``statistical noise'', whose standard deviation behaves as $\calO(1/\sqrt{n})$~\cite[Section III.A]{Bloch2016}.
\begin{remark}
Our use of $\V{\Qcoden,\Qinnn}$ is motivated by the following considerations. As a strong converse states that the optimal code rate has no dependency to the target constraint in the first asymptotics, it is clear that there is no strong converse for the value of $\delta$ \ac{wrt} the covert throughput, i.e., the covert throughput depends on the value $\delta$ through $d=\invQdelta$, which is directly related to $\V{\Qcoden,\Qinnn}$. This can be seen from Definition~\ref{def:cvt-tp-pair} and Theorem~\ref{thm:covert-capacity} where the notion of throughput depends on the covertness metric; hence, the choice of covertness metric matters. Many earlier works~\cite{Bloch2016,Wang2016a,Abdelaziz2017,Bendary2019,Bendary2020} measure covertness using the relative entropy $\avgD{\Qcoden}{\Qinnn}$. Unfortunately, relative entropy is only a loose proxy for variational distance since Pinsker's inequality is not tight~\cite{Tahmasbi2019} and is then less directly related to the operational test of the adversary. Furthermore, both $\avgD{\Qcoden}{\Qinnn}$ and $\avgD{\Qinnn}{\Qcoden}$ could in principle be used but, depending on which metric is chosen, different conclusions regarding the optimal signaling over \ac{AWGN} channels can be reached~\cite{Yan2019}.
\end{remark}
\begin{remark}
Our model does not include a power constraint on the channel input. This is justified since we only consider channel matrices with trivial null space and since any power constraint on the input is weaker than the covertness constraint~\cite[Section V.]{Wang2016a}.  Previous works \cite{Abdelaziz2017,Bendary2019,Bendary2020} impose the power constraint precisely because they allow non-trivial null spaces. 
\end{remark}

\section{Main Results}
\label{sec:main-results}

\begin{theorem}
  \label{thm:covert-capacity}
  The covert capacity of a \ac{MIMO}-\ac{AWGN} channel with full knowledge of the channel matrices is
  \begin{align}
  \label{eq:cvt-cap}
    C_{\textnormal{covert}} = \frac{\sigma_w^2}{\sigma_b^2}\sqrt{2\tr{\vct{\Lambda}_b^4\left(\vct{\Lambda}_w^{-1}\right)^4}}.
  \end{align}
  The covert capacity is achievable with key throughput
  \begin{align}
    \label{eq:key-tpt-achv}
    R_{\textnormal{key}}=&\sqrt{\frac{2}{{\tr{\vct{\Lambda}_b^4\left(\vct{\Lambda}_w^{-1}\right)^4}}}}\left(\tr{\vct{\Lambda}_b^2\left(\vct{\Lambda}_w^{-1}\right)^2}-\frac{\sigma_w^2}{\sigma_b^2}\tr{\vct{\Lambda}_b^4\left(\vct{\Lambda}_w^{-1}\right)^4}\right)^+,
  \end{align}
where $(x)^+\eqdef\max(x, 0)$.
\end{theorem}

\subsection{Converse Proof for Variational Distance}
\begin{proposition}
\label{prop:converse-V}
Consider a sequence of covert \ac{MIMO}-\ac{AWGN} communication schemes for the model in \eqref{eq:MIMO-scheme-def} with increasing blocklength $n\in\bbN^*$, characterized by $\epsilon_n\eqdef P_e^{(n)}$ and $\delta\geq\V{\Qcoden, \Qinnn}$. If $\lim_{n\to\infty} \epsilon_n = 0$ and $\limn{M_n}=\infty$, then we have 
\begin{equation}
    \liminf{\frac{\log M_n}{\sqrt{n}\invQfunc{\frac{1-\delta}{2}}}}\leq \frac{\sigma_w^2}{\sigma_b^2}\sqrt{2\tr{\vct{\Lambda}_b^4\left(\vct{\Lambda}_w^{-1}\right)^4}}.
\end{equation}
\end{proposition}
\begin{IEEEproof}
  The proof extends the techniques developed in \cite{Bash2013a,Tahmasbi2019,Zhang2019a} by constructing a test for Willie that is simple enough to be analyzed yet powerful enough to obtain a tight bound. Since we do not have any knowledge of the specific code exploited in the converse proof, the crux of our converse is to show that there cannot be too many high-power codewords, for otherwise the covertness would be compromised. We then analyze the maximal size of the low-power subcode, which is ``good'' in the sense that both reliability and covertness can be ensured. We first recall the Berry-Esseen Theorem.
\begin{theorem}[Berry-Esseen Theorem]
\label[theorem]{thm:berry-esseen}
Let $X_1,\dotsc,X_n$ be independent random variables such that for $k\in\intseq{1}{n}$, we have $\expect{X_k}=\mu_k$, $\sigma_k^2=\Var{X_k}$, and $t_k=\expect{|X_k-\mu_k|^3}$. If we define $\sigma^2=\sum_{k=1}^n\sigma_k^2$ and $T=\sum_{k=1}^nt_k$, then we have
\begin{align}
    \left|\P{\sum_{k=1}^n(X_k-\mu_k)\geq\lambda\sigma}-\Qfunc{\lambda}\right|\leq\frac{6T}{\sigma^3}.
\end{align}
\end{theorem}
\paragraph{Lower bound on covertness metric}
We start by establishing a lower bound relating the covertness metric to the minimum received power of codewords at Bob within a given code $\calM$.
Consider a simple hypothesis testing problem with two hypotheses $H_0$ and $H_1$ corresponding to distributions $\Qinnn$ and $\Qcoden$, respectively. We define a sub-optimal power detector  
\begin{equation}
    T\left(\vct{z}^n\right)\eqdef\indic{\sum_{i=1}^nS_i > \tau},
\end{equation}
where $S_i\eqdef S\left(\vct{z}_i\right)\eqdef\norm[2]{\smash{\vct{H}_b\left(\vct{H}_w\right)^\dagger\vct{z}_i}}^2$, and the threshold $\tau$ will be specified later. The intuition behind the test is to realign Willie's observations with those of Bob. Note that, $\vct{H}_w^\hermittian\vct{H}_w$ is invertible because of the full-rank assumption. Hence, we rewrite the test $S_i$ using the \ac{GSVD} as 
\begin{equation}
    S_i=\left(\vct{H}_b\left(\vct{H}_w\right)^\dagger\vct{z}_i\right)^\hermittian\left(\vct{H}_b\left(\vct{H}_w\right)^\dagger\vct{z}_i\right)=\hat{\vct{z}}_i^\hermittian\hat{\vct{z}}_i,
\end{equation}
where $\hat{\vct{z}}_i=\vct{\Lambda}_b\vct{\Lambda}_w^{-1}\vct{U}_w^\hermittian\vct{z}_i$. The following lemma, which is proved in Appendix~\ref{sec:pf-lm-code-min-power-bound}, characterizes upper bounds for both the false-alarm and the missed-detection probabilities.
\begin{lemma}
  \label{lemma:code-min-power-bound}
Consider a specific code $\calM$ with codewords indexed by $k$, $\vct{x}^{\left(k\right)n}=\matvct{\vct{x}^{\left(k\right)}}{1}{n}\in\calC$. By defining $P_*\eqdef \min_{k}\norm[F]{\vct{H}_b\vct{x}^{(k)n}}^2=\min_{k}\tr{\vct{\Lambda}_b^2\vct{P}^{(k)}}$ the minimum power of Bob's received codewords, and setting the detection threshold to $\tau=\frac{P_*}{2}+n\sigma_w^2\tr{\vct{\Lambda}_b^2\left(\vct{\Lambda}_w^{-1}\right)^2}$,
\begin{align}
  \label{eq:V-conv-alpha}
    \alpha&\leq\Qfunc{\frac{P_*}{2\sqrt{2n \tr{\vct{\Lambda}_b^4\left(\vct{\Lambda}_w^{-1}\right)^4} }\sigma_w^2}}+\frac{B_0}{\sqrt{n}},\\
    \beta &\leq\Qfunc{\frac{P_*}{2\sqrt{2n\tr{\vct{\Lambda}_b^4\left(\vct{\Lambda}_w^{-1}\right)^4}}\sigma_w^2}}+\frac{P_*^2\tr{\vct{\Lambda}_b^2\left(\vct{\Lambda}_w^{-1}\right)^2}}{4\sqrt{\pi}n^{3/2}\tr{\vct{\Lambda}_b^4\left(\vct{\Lambda}_w^{-1}\right)^4}^{3/2}\sigma_w^4} + \frac{B_1}{\sqrt{n}},
\end{align}
where $B_0$ and $B_1$ are some constants independent of $n$.
\end{lemma}

Hence, the covertness metric can be lower-bounded as
\begin{align}
    \label{eq:V-converse-apprx}
    &\V{\Qcoden,\Qinnn}\geq1-\alpha-\beta \nonumber\\
            &\geq1-2\Qfunc{\frac{P_*}{2\sqrt{2n\tr{\vct{\Lambda}_b^4\left(\vct{\Lambda}_w^{-1}\right)^4}}\sigma_w^2}}-\frac{P_*^2\tr{\vct{\Lambda}_b^2\left(\vct{\Lambda}_w^{-1}\right)^2}}{4\sqrt{\pi}n^{3/2}\tr{\vct{\Lambda}_b^4\left(\vct{\Lambda}_w^{-1}\right)^4}^{3/2}\sigma_w^4}-\frac{B_0+B_1}{\sqrt{n}},
\end{align}
which only depends on the code through the minimum power of Bob's received codewords.
\paragraph{Existence of a good sub-code} For a covert code $\calC$, we develop a bound for the maximum power of a non-empty low-power sub-code in the following lemma, which is proved in Appendix~\ref{sec:proof-lm-V-conv-sub-code-power}. The key idea is to use \eqref{eq:V-converse-apprx} to analyze the covertness for the high-power sub-code and argue the existence of a low-power sub-code.
\begin{lemma}
\label{lemma:V-conv-sub-code-power}
For any covert channel code $\calC$, given a decreasing sequence $\{\gamma_n\}_{n=1}^\infty$ with $\gamma_n\in (0,1)$, $\limn{\gamma_n}=0$, 
there exists a subset of codewords $\calC^{(\ell)}$ such that $\card{\calC^{(\ell)}}\geq\gamma_n\card{\calC}$ and 
$\norm[F]{\vct{H}_b\vct{x}^n}^2\leq A\sqrt{n},$
where \begin{align}
    \label{eq:maximum-power-of-codebook}
        A&\eqdef2\sqrt{2\tr{\vct{\Lambda}_b^4\left(\vct{\Lambda}_w^{-1}\right)^4}}\sigma_w^2 \invQfunc{\frac{1-\delta}{2}-\frac{\nu^2\tr{\vct{\Lambda}_b^2\left(\vct{\Lambda}_w^{-1}\right)^2}}{4\sqrt{\pi n}\tr{\vct{\Lambda}_b^4\left(\vct{\Lambda}_w^{-1}\right)^4}^{3/2}\sigma_w^4}-\gamma_n},
\end{align} and $\nu$ depends on the channel.
\end{lemma}

\paragraph{Upper bound on covert message size within a good sub-code}
\textcolor{black}{The code $\calC$ can be partitioned into $K_n$ sub-codes $\calC_s$ indexed by the key value $s$ for all $s\in\intseq{1}{K_n}$ such that $\calC=\cup_{s\in\intseq{1}{K_n}}\calC_s$, and the size of each sub-code is $M_n$.} Let $\calC^{\left(\ell\right)}_s\eqdef\calC_s\cap\calC^{\left(\ell\right)}$. By the pigeonhole principle, there exists a sub-code $\calC_s$ satisfying $\card{\calC_s^{\left(\ell\right)}}\geq\gamma_n M_n$. Furthermore, since the average probability of error of $\calC_{s}$ is at most $\epsilon_n$, we have  $\bar{P}^{(n)}_e\left(\calC_s^{\left(\ell\right)}\right)\leq\frac{\epsilon_n}{\gamma_n}$, which vanishes in the limit of large $n$ upon choosing $\{\gamma_n\}_{n=1}^\infty$ such that $\limn{\frac{\epsilon_n}{\gamma_n}}=0$. 

Let $\widetilde{W}$ denote the uniformly distributed variable over the messages in $\calC_s^{\left(\ell\right)}$. By standard techniques, we therefore have
\begin{align}
  \log\card{\calC_s^{\left(\ell\right)}}&=\avgH{\widetilde{W}\vert S=s}\\&
  = \avgI{\widetilde{W};\vct{Y}^n|S=s} + \avgH{\widetilde{W}|\vct{Y}^nS=s}\\
    &\leq \avgI{\widetilde{W};\vct{Y}^n|S=s} + \left[\frac{\epsilon_n}{\gamma_n}\log\card{\calC_s^{\left(\ell\right)}}+\Hb{\frac{\epsilon_n}{\gamma_n}}\right]\\
    &\leq \avgI{\vct{X}^n;\vct{Y}^n|S=s} + \left[\frac{\epsilon_n}{\gamma_n}\log\card{\calC_s^{\left(\ell\right)}}+\Hb{\frac{\epsilon_n}{\gamma_n}}\right]\\
    \label{eq:V-card-I}
                                        &\leq n\avgI{\bar{\vct{X}};\bar{\vct{Y}}} + \frac{\epsilon_n}{\gamma_n}\log\card{\calC_s^{\left(\ell\right)}}+1,
\end{align}
where the random variables $\bar{\vct{X}}$ and $\bar{\vct{Y}}$ have distributions
\begin{align*}
\Pi_{\bar{\vct{X}}}\left(\vct{x}\right)&\eqdef\frac{1}{n}\sum_{i=1}^n\Pi_{\vct{X}_i}\left(\vct{x}\right)=\frac{1}{n}\sum_{i=1}^n\frac{1}{\card{\calC_s^{\left(\ell\right)}}}\sum_{\vct{x}^n\in\calC_s^{\left(\ell\right)}}\indic{\vct{x}=\vct{x}_i}\\
  ~\text{and}~P_{\bar{\vct{X}}\bar{\vct{Y}}}&\eqdef\Pi_{\bar{\vct{X}}}\Wb.
\end{align*}
Let $\expect{\bar{\vct{X}}\bar{\vct{X}}^\hermittian}=\vct{Q}_n$. Note that $\expect{\bar{\vct{Y}}\bar{\vct{Y}}^\hermittian}=\vct{H}_b\vct{Q}_n\vct{H}_b^\hermittian+\sigma_b^2\vct{I}_{N_b}$. Then,  
\begin{align}
    \avgI{\bar{\vct{X}};\bar{\vct{Y}}}&=\h{\bar{\vct{Y}}}-\h{\bar{\vct{Y}}|\bar{\vct{X}}}\\
    & \leq\frac{1}{2}\log\det{\vct{I}_{N_b}+\frac{1}{\sigma_b^2}\vct{H}_b\vct{Q}_n\vct{H}_b^\hermittian}\\
    &=\frac{1}{2}\tr{\blog{\vct{I}_{N_b}+\frac{1}{\sigma_b^2}\vct{H}_b\vct{Q}_n\vct{H}_b^\hermittian}}\\
    &\labrel{\leq}{eq:V-conv-tr-log-ineq}\frac{1}{2\sigma_b^2}\tr{\vct{H}_b\vct{Q}_n\vct{H}_b^\hermittian}
    \label{eq:V-conv-I-ub}
    \labrel{\leq}{eq:V-conv-low-power} \frac{A}{2\sigma_b^2\sqrt{n}},
\end{align}
where \eqref{eq:V-conv-tr-log-ineq} follows since for any $\vct{A}\succcurlyeq\vct{0}\text{~and~}\norm[2]{\vct{A}}\leq 1$, $\tr{\blog{\vct{I+A}}}=\sum_{i}\blog{1+\lambda_i(\vct{A})}\leq\sum_{i}\lambda_i(\vct{A})=\tr{\vct{A}}$, where $\{\lambda_i(\vct{A})\}_i$ is the set of eigenvalues of $\vct{A}$ and we have used $\blog{1+x}\leq x$ for all $x>0$, and \eqref{eq:V-conv-low-power} follows from the definition of $\calC^{(\ell)}$ and  $\tr{\vct{H}_b\vct{Q}_n\vct{H}_b^\hermittian}=\frac{1}{n\card{\calC_s^{(\ell)}}}\sum_{\vct{x}^n\in\calC_s^{(\ell)}}\norm[F]{\vct{H}_b\vct{x}^n}^2$.
Combining~\eqref{eq:maximum-power-of-codebook}, \eqref{eq:V-card-I}, \eqref{eq:V-conv-I-ub}, and the fact that $\limn{\gamma_n}=0$, we have
\begin{equation}
\label{eq:conv-V-low-power-card-bound}
    \log\card{\calC_s^{\left(\ell\right)}}\leq\frac{\sqrt{2n\tr{\vct{\Lambda}_b^4\left(\vct{\Lambda}_w^{-1}\right)^4}}\frac{\sigma_w^2}{\sigma_b^2}\invQfunc{\frac{1-\delta}{2}}+\calO\left(1\right)}{1-\frac{\epsilon_n}{\gamma_n}}.
\end{equation}

We further choose the sequence $\{\gamma_n\}_{n=1}^\infty$ such that $\limn{-\frac{\log\gamma_n}{\sqrt{n}\invQdelta}}=0$. 
Finally, we obtain
\begin{align}
    \liminf{\frac{\log M_n}{\sqrt{n}\invQfunc{\frac{1-\delta}{2}}}}&\leq\liminf{\frac{\log\card{\calC_s^{\left(\ell\right)}}-\log\gamma_n}{\sqrt{n}\invQfunc{\frac{1-\delta}{2}}}}\\
    &=\frac{\sigma_w^2}{\sigma_b^2}\sqrt{2\tr{\vct{\Lambda}_b^4\left(\vct{\Lambda}_w^{-1}\right)^4}}.
\end{align}
\end{IEEEproof}
Unfortunately, we have not found a matching converse argument for the key throughput. 
\subsection{Achievability Proof for Variational Distance}
\label{subsec:V-achv}
\begin{proposition}
\label{prop:achv-V}
Consider a \ac{MIMO}-\ac{AWGN} covert communication channel in \eqref{eq:MIMO-scheme-def}. 
There exist covert communication schemes such that
\begin{align}
    \label{eq:achv-V-message-const}
  &\limn{\frac{\log M_n}{\sqrt{n}\invQfunc{\frac{1-\delta}{2}}}}
  \geq\frac{\sigma_w^2}{\sigma_b^2}\sqrt{2\tr{\vct{\Lambda}_b^4\left(\vct{\Lambda}_w^{-1}\right)^4}}, \\
    \label{eq:achv-V-rsl-code-const}
  &\limn{\frac{\log M_nK_n}{\sqrt{n}\invQfunc{\frac{1-\delta}{2}}}} 
  \leq\sqrt{\frac{2}{\tr{\vct{\Lambda}_b^4\left(\vct{\Lambda}_w^{-1}\right)^4}}}\tr{\vct{\Lambda}_b^2\left(\vct{\Lambda}_w^{-1}\right)^2},\\
    &\limn{P_e^{(n)}}=0,~\V{\Qcoden,\Qinnn}\leq\delta.
\end{align}
\end{proposition}

\begin{IEEEproof}
Our proof follows \cite{Wang2016a,Bloch2016,Zhang2019a} to construct a \ac{BPSK} code achieving the desired throughput pair. Note that we could also use a Gaussian codebook, but this would require extra care to deal with the power of codewords.
\paragraph{Covert stochastic process~\cite{Bloch2016}}
We introduce another input process $\PiQn$ with covariance matrix $\vct{Q}_n$ and its associated distribution at the output of channel $\Ww$,
    $\Qinf\eqdef\PiQn\circ\Ww.$
Additionally, the associated product distributions are
    $\PiQnn=\prod_{i=1}^n\PiQn ~\text{and}~\Qinfn=\prod_{i=1}^n\Qinf.$
The achievability proof decomposes the covertness metric $\V{\Qcoden,\Qinnn}$ into two pieces, $\V{\Qcoden,\Qinfn}$ and $\V{\Qinfn, \Qinnn}$ by the triangle inequality. The former term is related to the channel output approximation problem, and we rely on the channel resolvability to analyze its behavior~\cite{Bloch2016,Han1993}. We upper-bound the latter term by a covertness constraint $\delta-\frac{1}{\sqrt{n}}$. Essentially, this constraint makes $\Qinfn$ asymptotically indistinguishable from the output distribution of the innocent symbol $\Qinnn$; accordingly, $\Qinfn$ is called a \emph{covert stochastic process}. The rationale for introducing such a process is to find a proxy to control the discrepancy captured by the covertness metrics by a carefully designed covariance matrix $\vct{Q}_n$, which is the counterpart of low-weight codewords designed in the covert communication scheme over DMCs~\cite{Bloch2016,Tahmasbi2019}.

\paragraph{Random code generation}
We decompose the channel into $m$ parallel sub-channels defined by the \ac{GSVD} precoding with the input alphabet $\widetilde{\calX}\eqdef\{-a_{n,1}, 0, a_{n,1}\}\times\dotsc\times\{-a_{n,m}, 0, a_{n,m}\}$, where $m=\rk{\vct{H}_b}=\rk{\vct{H}_w}=N_a$. Throughout the section, tildes refer to the operations over the parallel sub-channels. Let $M_n, K_n\in\bbN^*.$ 
Alice independently generates $M_nK_n$ codewords $\tilde{\vct{x}}^n\left(\ell, k\right)\in\prod_{j=1}^m\{-a_{n,j},a_{n,j}\}^{n}$ jointly over all the sub-channels with $\ell\in\intseq{1}{M_n}$ and $k\in\intseq{1}{K_n}$, according to the distribution $\prod_{j=1}^m\Pi_{\rho_{n,j}}$ such that 
    $\Pi_{\rho_{n,j}}\left(a_{n,j}\right)=\Pi_{\rho_{n,j}}\left(-a_{n,j}\right)=\frac{1}{2},~\text{and}~\Pi_{\rho_{n,j}}\left(0\right)=0,$
where $\{\rho_{n,j}\}$ is a set of non-negative real numbers defined as
\begin{equation}
\label{eq:achv-V-rho-def}
    \rho_{n,j}\eqdef\frac{\tau_j\invQfunc{\frac{1-\delta}{2}}}{\sqrt{n}}=a_{n,j}^2,~\forall j\in\intseq{1}{m},
\end{equation}
and $\{\tau_j\}_{j=1}^m$ is determined later via an optimization program. 
We define two diagonal matrices $\vct{P}_n$ and $\vct{T}$ with $\{\rho_{n,j}\}_{j=1}^m$ and $\{\tau_j\}_{j=1}^m$ as the diagonal entries, respectively. 
For simplicity, we stack codewords into $\tilde{\vct{x}}^n\left(\ell,k\right)\in\bbR^{m\times n}$. 
Alice then employs the precoding matrix $(\vct{V}^\hermittian)^{-1}$ to form $\vct{x}^n=\left(\vct{V}^\hermittian\right)^{-1}\parVarx^n$, and therefore the input covariance matrix after the precoding is $\vct{Q}_n=(\vct{V}^\hermittian)^{-1}\vct{P}_n\vct{V}^{-1}$, where we design $\vct{P}_n$ carefully as in \eqref{eq:achv-V-rho-def}. Bob and Willie postprocess their observations from channel outputs $\vct{y}^n\in\bbR^{N_b\times n}~\text{and~}\vct{z}^n\in\bbR^{N_w\times n}$ by transforming them via $\vct{U}_b^\hermittian~\text{and}~\vct{U}_w^\hermittian$ to get $\tilde{\vct{y}}^n\in\bbR^{m\times n}~\text{and}~\tilde{\vct{z}}^n\in\bbR^{m\times n}$, respectively. \textcolor{black}{There is no loss of generality in making this assumption for Willie, as the post-processing $\vct{U}_w^\hermittian$ performs an orthogonal transform and then discards the components in the observations corresponding to $\text{Null}(\vct{H}_w^\hermittian)$, which only contain noise.} These operations result in, $\forall i\in\intseq{1}{n}$, 
    $\tilde{\vct{y}}_i=\vct{\Lambda}_b\tilde{\vct{x}}_i+\tilde{\vct{n}}_{b,i},~\text{and}~\tilde{\vct{z}}_i=\vct{\Lambda}_w\tilde{\vct{x}}_i+\tilde{\vct{n}}_{w,i},$
where $\tilde{\vct{n}}_b\sim\N{\vct{0}}{\sigma_b^2\vct{I}_m}$ and $\tilde{\vct{n}}_w\sim\N{\vct{0}}{\sigma_w^2\vct{I}_m}$.
From the perspective of the $m$ sub-channels $\left(\widetilde{\calX},\parWb,\widetilde{\calY}, \parWw,\widetilde{\calZ}\right)$ with $\widetilde{\calY}=\bbR^m,~\text{and}~\widetilde{\calZ}=\bbR^m$, we therefore define the following statistics:
    $\PiPn\left(\tilde{\vct{x}}\right)=\prod_{j=1}^m\Pi_{\rho_{n,j}}\left(\tilde{x}_j\right),$ $\PiPnn=\prod_{i=1}^n\PiPn,$
$\parPinf=\PiPn\circ\parWb$, $\parPinfn=\prod_{i=1}^n \parPinf, $
    $\parQinf=\PiPn\circ\parWw$, and $\parQinfn=\prod_{i=1}^n\parQinf.$
Note that because of the parallelness of sub-channels, we can derive simple forms as follows:
\begin{align}
    \parPinf&=\prod_{j=1}^m\left(\frac{1}{2}\N{-\lambda_{b,j}a_{n,j}}{\sigma_b^2}+\frac{1}{2}\N{\lambda_{b,j}a_{n,j}}{\sigma_b^2}\right),\\
    \parQinf&=\prod_{j=1}^m\left(\frac{1}{2}\N{-\lambda_{w,j}a_{n,j}}{\sigma_w^2}+\frac{1}{2}\N{\lambda_{w,j}a_{n,j}}{\sigma_w^2}\right).
\end{align}
In the sequel, we also use the notation $\parjPinf\eqdef\frac{1}{2}\N{-\lambda_{b,j}a_{n,j}}{\sigma_b^2}+\frac{1}{2}\N{\lambda_{b,j}a_{n,j}}{\sigma_b^2}$ and $\parjQinf\eqdef\frac{1}{2}\N{-\lambda_{w,j}a_{n,j}}{\sigma_w^2}+\frac{1}{2}\N{\lambda_{w,j}a_{n,j}}{\sigma_w^2}$ to represent distributions at each sub-channel in the above decomposition. Similarly, we also use $\parjPinn\eqdef\N{0}{\sigma_b^2}$ and $\parjQinn\eqdef\N{0}{\sigma_w^2}$. Hence, $\parPinn=\prod_{j=1}^m\parjPinn$ and $\parQinn=\prod_{j=1}^m\parjQinn$.
\paragraph{Channel reliability analysis}

\begin{lemma}
\label[lemma]{lemma:achv-V-rel}
By choosing 
\begin{equation}
\label{eq:achv-V-msg-size}
    \log M_{n}= (1-\xi)\frac{\sqrt{n}\invQdelta}{2\sigma_b^2}\tr{\vct{\Lambda}_b^2\vct{T}},
\end{equation}
the average probability of error satisfies
\begin{align}
    \expect{\bar{P}^{(n)}_{e}}\leq e^{-\theta_{1}\sqrt{n}\invQdelta},
\end{align}
where $\xi\in(0,1)$, and $\theta_{1}>0.$
\end{lemma}
The proof is provided in Appendix~\ref{sec:proof-lm-achv-V-rel}.

\paragraph{Covertness analysis}
\begin{lemma}
\label[lemma]{lemma:achv-V-rsl}
By choosing $\vct{T}$ such that
\begin{align}
    \label{eq:achv-V-ch-rsl-cvt-cnst}
    \frac{1}{4\sigma_w^4}\tr{\vct{\Lambda}_w^4\vct{T}^2}\leq 2 - \frac{C}{\sqrt{n}\invQdelta}, 
\end{align}
for some $C>0$, and 
\begin{equation}
    \label{eq:achv-V-ch-rsl-code-size}
    \log M_nK_n=\left(1+\xi\right)\frac{\sqrt{n}\invQdelta}{2\sigma_w^2}\tr{\vct{\Lambda}_w^2\vct{T}},
\end{equation}
the expected covertness metric is bounded as follows:
\begin{equation}
\label{eq:V-close-delta}
    \expect{\V{\Qcoden,\Qinnn}}\leq \delta+ e^{-\theta_2\sqrt{n}\invQdelta}-\frac{1}{\sqrt{n}},
\end{equation}
where $\xi\in(0,1)$, and $\theta_{2}>0$ are some constants.
\end{lemma}
The proof is provided in Appendix~\ref{sec:proof-lm-achv-V-rsl}.

\paragraph{Identification of a specific code}
Choosing $\xi$, $\log M_{n}$ and $\log K_n$ to satisfy \Cref{lemma:achv-V-rel} and \Cref{lemma:achv-V-rsl}, Markov's inequality allows us to conclude that there exists at least one specific code $\calC$ with $n$ large enough and appropriate constants $\xi_{1}, \xi_2>0$ such that 
    $\bar{P}^{(n)}_{e}\leq e^{-\xi_{1}\sqrt{n}\invQdelta}$ and $\V{\Qcoden,\Qinnn}\leq \delta-\frac{1}{\sqrt{n}} +e^{-\xi_2\sqrt{n}\invQdelta}.$
    Although a code $\calC$ with vanishing $\bar{P}^{(n)}_e$ does not necessary satisfy the reliability constraint~\eqref{eq:max-P-err}, which requires $P_e^{(n)}$ to vanish as $n$ goes to infinity, the following lemma from~\cite{Zhang2019a} gives us such a guarantee by merely rearranging the codewords in $\calC$.
\begin{lemma}
\label[lemma]{lemma:codeword-rearrange}
Suppose a code $\calC$ contains $K_n$ sub-codes of size $M_n$ such that $\bar{P}^{(n)}_e\leq\epsilon_n$ and $\V{\Qcoden,\Qinnn}\leq e^{-\xi_2\sqrt{n}\invQdelta}+\delta-\frac{1}{\sqrt{n}}$. Then, there exists a code $\calC^\prime$ containing $K_n^\prime$ sub-codes of size $M_n^\prime$ such that $P^{(n)}_e\leq\epsilon_n^\prime$ and $\V{\Qcoden,\Qinnn}\leq e^{-\xi_2\sqrt{n}\invQdelta}+\delta-\frac{1}{\sqrt{n}}$. In particular, $\limn{\epsilon_n}=\limn{\epsilon_n^\prime}=0$, $\limn{\frac{M_n^\prime}{M_n}}=1$, and $\limn{\frac{K_n^\prime}{K_n}}=1.$
\end{lemma}
\paragraph{Constellation power design}
\label{par:V-achv-const-power}
Next, we design the optimal constellation points that result in the largest achievable message set size satisfying the covertness constraint. We formalize our optimization program by combining~\eqref{eq:achv-V-msg-size} and \eqref{eq:achv-V-ch-rsl-cvt-cnst} as follows:
\begin{subequations}
\label{eq:V-opt-1}
    \begin{align}
    \label{eq:V-opt-1-obj}
        &\max_{\vct{T}\succcurlyeq\vct{0}} ~\frac{1}{2\sigma_b^2}\tr{\vct{\Lambda}_b^2\vct{T}}, \\
    \label{eq:V-opt-1-con}
        &\text{s.t.~} \frac{1}{4\sigma_w^4}\tr{\vct{\Lambda}_w^4\vct{T}^2} \leq 2 - \frac{C}{\sqrt{n}\invQdelta}.
    \end{align}
\end{subequations}
To solve this, we regard the term $\frac{C}{\sqrt{n}\invQdelta}$ as a perturbation. Consider the optimization
\begin{subequations}
\label{eq:V-opt-2}
    \begin{align}
    \label{eq:V-opt-2-obj}
        &\max_{\vct{T}\succcurlyeq\vct{0}} ~\frac{1}{2\sigma_b^2}\tr{\vct{\Lambda}_b^2\vct{T}}, \\
    \label{eq:V-opt-2-con}
        &\text{s.t.~} \frac{1}{4\sigma_w^4}\tr{\vct{\Lambda}_w^4\vct{T}^2} \leq 2.
    \end{align}
\end{subequations}
The optimal Lagrange multiplier $\mu$ and solution $\vct{T}$ to \eqref{eq:V-opt-2} are
    $\mu = \frac{\sigma_w^2}{2\sqrt{2}\sigma_b^2}\sqrt{\tr{\vct{\Lambda}_b^4\left(\vct{\Lambda}_w^{-1}\right)^4}}$, and
   $\vct{T}=2\sqrt{2}\sigma_w^2\frac{\vct{\Lambda}_b^2\left(\vct{\Lambda}_w^{-1}\right)^4}{\sqrt{\tr{\vct{\Lambda}_b^4\left(\vct{\Lambda}_w^{-1}\right)^{4}}}},$
respectively.
Let $\rho$ and $\rho^\prime$ denote the optimal objective values of \eqref{eq:V-opt-2} and \eqref{eq:V-opt-1}, respectively. By the sensitivity analysis~\cite[Ch. 8.5]{Luenberger1997optbyvec}, we have 
\begin{equation}
  \label{eq:V-opt-2-perturbation}
    \rho\geq\rho^\prime\geq\rho-\calO\left(\frac{1}{\sqrt{n}\invQdelta}\right),
\end{equation}
which shows the perturbation is negligible as $n$ goes to infinity.
Consequently, 
\begin{align}
    &\limn{\frac{\log M_n}{\sqrt{n}\invQdelta}}=\left(1-\xi\right)\frac{\sigma_w^2}{\sigma_b^2}\sqrt{2\tr{\vct{\Lambda}_b^4\left(\vct{\Lambda}_w^{-1}\right)^4}},\\
        \label{eq:V-achv-key-msg-tpt}
        &\limn{\frac{\log M_nK_n}{\sqrt{n}\invQdelta}}=\left(1+\xi\right)\frac{\sqrt{2}\tr{\vct{\Lambda}_b^2\left(\vct{\Lambda}_w^{-1}\right)^2}}{\sqrt{\tr{\vct{\Lambda}_b^4\left(\vct{\Lambda}_w^{-1}\right)^4}}}.
\end{align}
Since for any $\xi\in(0,1)$, there exists a scheme satisfying all the requirements, we can therefore make $\xi$ arbitrarily small, i.e., $\xi\to 0^+$. The result then follows.

The reader might wonder why the optimal solution to \eqref{eq:V-opt-2} is not the usual water-filling solution. This is a unique phenomenon due to the covertness constraint. The usual water-filling solution would encourage the use of high power in the sub-channels in which Bob has better observations than Willie. In contrast, the square-root law discourages the use of high power, as allocating too much power to sub-channels that have better observations increases the risk of detection. Hence, one should not expect the water-filling solution to appear here. Specifically, our power allocation uses all the sub-channels and suggests that each sub-channel $j$ contribute $\left(1+\xi\right)\frac{\sqrt{2}\lambda_{b,j}^2\lambda_{w,j}^{-2}}{\sqrt{\tr{\vct{\Lambda}_b^4\left(\vct{\Lambda}_w^{-1}\right)^4}}}\sqrt{n}\invQdelta$ to $\log M_nK_n$, which is aligned with the direction of the diagonal elements of $\left(\vct{\Lambda}_b^2\left(\vct{\Lambda}_w^{-1}\right)^2\right)$.
\end{IEEEproof}

\section{Covert Communication with Unknown Warden Channel State}
We now assume that only partial channel state information of Willie's channel is available. Specifically, all parties know the exact $\vct{H}_b$, while Alice only knows that $\vct{H}_w$ belongs to the following uncertainty set:
\begin{align}
\label{eq:uncertainty-set-def}
  \calS\eqdef\{\vct{H}_w=\vct{U}_w\vct{\Lambda}_w\vct{V}^\hermittian:\norm[2]{\vct{\Lambda}_w}\leq\lambda_0, \nonumber\\
  m=\rk{\vct{H}_w}=\rk{\vct{H}_b}=N_a\},
\end{align}
where $\vct{U}_w$ is known to Willie and hence can be canceled by post-processing~\cite{Schaefer2015}. Thus, the set $\calS$ contains all the channels that are fully aligned with the main channel and for which the singular-value matrix is less than or equal to $\vct{\Lambda}_0\eqdef\lambda_0\vct{I}_m$. The channel realization is fixed during the transmission period. \textcolor{black}{This model corresponds to a quasi-static scenario where the adversary cannot be closer to the transmitter than a certain protection distance~\cite{Schaefer2015}.}

\sloppy For an $\left(n,M_n,K_n,\delta\right)$-code $\calC$ designed for the compound channel induced by $\calS$, the covertness metric at Willie is
    $\sup_{\vct{H}_w\in\calS}\V{\Qcoden_{\vct{H}_w},\Qinnn}$,
where $\Qcoden_{\vct{H}_w}$ is the distribution when communication occurs over the channel realization $\vct{H}_w$, 
\begin{align}
\Qcoden_{\vct{H}_w}\left(\vct{z}^n\right)=\frac{1}{M_nK_n}\sum_{\ell=1}^{M_n}\sum_{k=1}^{K_n}\Wwn\left(\vct{z}^n|\vct{x}^{(\ell k)n}\right),
\end{align}
$\forall \vct{z}^n\in\bbR^{N_w\times n},$ and $W_{\vct{Z}|\vct{X}=\vct{x}}\sim\N{\vct{H}_w\vct{x}}{\sigma_w^2\vct{I}_{N_w}}$. 

We show that the compound covert capacity is equal to the worst-case covert capacity at channel realization $\vct{U}_w\vct{\Lambda}_0\vct{V}^\intercal$. Here we only present the achievability proof for the compound covert capacity under the variational distance, in which we show that there exists a compound covert code achieving the worst-case covert capacity. The converse proof follows from the fact that the worst-case covert capacity within the uncertainty set $\calS$ upper-bounds the compound covert capacity, for a compound covert code also works on the worst-case channel realization by definition as pointed out in \cite[Corollary 1]{Schaefer2015}.
\begin{proposition}
\label{prop:compound-ach}
Consider a compound \ac{MIMO}-\ac{AWGN} covert communication channel in \eqref{eq:MIMO-scheme-def} and the uncertainty set $\calS$ in~\eqref{eq:uncertainty-set-def} containing all possible channel realizations of the warden. 
There exist covert communication schemes such that
\begin{align}
  \label{eq:compound-opt-tp-stat}
  &\limn{\frac{\log M_n}{\sqrt{n}\invQfunc{\frac{1-\delta}{2}}}}\geq\frac{\sigma_w^2}{\sigma_b^2}\sqrt{2\tr{\vct{\Lambda}_b^4\left(\vct{\Lambda}_0^{-1}\right)^4}}, \\
  \label{eq:compound-opt-k-stat}
  &\limn{\frac{\log M_nK_n}{\sqrt{n}\invQfunc{\frac{1-\delta}{2}}}}\leq\sqrt{\frac{2}{\tr{\vct{\Lambda}_b^4}}}\tr{\vct{\Lambda}_b^2},\\
    &\limn{P_e}=0,~\sup_{\vct{H}_w\in\calS}\V{\Qcoden_{\vct{H}_w},\Qinnn}\leq\delta.
\end{align}
\end{proposition}
Note that \eqref{eq:compound-opt-k-stat} does not depend on Willie's channel. As mentioned previously, the power allocation makes each sub-channel $j$ contribute $\left(1+\xi\right)\frac{\sqrt{2}\lambda_{b,j}^2\lambda_{0}^{-2}}{\sqrt{\tr{\vct{\Lambda}_b^4\left(\vct{\Lambda}_0^{-1}\right)^4}}}\sqrt{n}\invQdelta$ to $\log M_nK_n$, which is aligned with the direction of the diagonal elements of $\left(\vct{\Lambda}_b^2\left(\vct{\Lambda}_0^{-1}\right)^2\right)$. Since we show that the worst-case channel capacity at $\vct{\Lambda}_0$ is achievable, the fact that $\vct{\Lambda}_0$ is isotropic makes the result independent of Willie's channel.
\begin{IEEEproof}
We extends the proof in~\cite{Schaefer2015}, using ideas from \cite{He2014}. The idea of the proof in \cite{Schaefer2015} is to extend the result of compound secrecy capacity for uncountably infinite compound \acp{DMC} to continuous alphabets through a sequence of successively finer quantizers, which quantize the input and output alphabets at all parties. The compound secrecy rate derived from quantized alphabets can be made arbitrarily close to the compound secrecy capacity with a sufficiently fine quantizer. Unfortunately, this process requires the adversary to obey the quantization rule and implicitly assumes that the adversary should cooperate with Alice and Bob. We propose a small correction that circumvents the issue by considering an adversary that directly operates on the channel output without quantization, and directly analyzes the difference in terms of covertness induced by a code between two close channel states.
\paragraph{Discretization} Since the uncertainty set $\calS$ described in \eqref{eq:uncertainty-set-def} is uncountable, we first discretize $\calS$ to construct a countably finite uncertainty set $\calS_n$ with a suitable choice of discretization level and discretization points.

Note that since the uncertainty set $\calS$ is subject to the spectral norm constraint, which results in an $m$-dimensional hypercube with length $\lambda_0$ on each side, a natural way to discretize is to uniformly slice $\calS$ into $2^{mn}$ hypercubic regions with length $\epsilon_n\eqdef\lambda_0 2^{-n}$ on each side. The discretization points constructing the set $\calS_n$ are chosen as follows:
\begin{align}
    \calS_n\eqdef&\{\vct{H}=\vct{U}\vct{\Lambda}_{J}\vct{V}^\hermittian:\vct{\Lambda}_J=\diag{j_1\epsilon_n,\dotsc,j_m\epsilon_n}, \nonumber\\
    &J=\left(j_1,\dotsc,j_m\right), j_\ell\in\intseq{1}{2^n},\forall\ell\in\intseq{1}{m}\},
\end{align}
where $J$ is an index for the elements in $\calS_n$, and since $\vct{U}$ is known to Willie, we henceforth omit its impact in the remaining.

Each discretization point $\vct{H}_J$ is associated with a neighborhood 
\begin{align}
    \calS_{J,n}\eqdef\{\widetilde{\vct{H}}=\vct{U}\widetilde{\vct{\Lambda}}\vct{V}^\hermittian:\vct{\Lambda}_J\succcurlyeq\widetilde{\vct{\Lambda}}, \norm[2]{\smash{\vct{\Lambda}_J-\widetilde{\vct{\Lambda}}}}<\epsilon_n\},
\end{align}
which covers a portion of the original uncertainty set. By construction, $\cup_{\vct{H}_J\in\calS_n}\calS_{J,n}=\calS$.
As discussed in previous sections, without loss of generality, we directly investigate the parallel sub-channels described by $\vct{\Lambda}_b$ and $\vct{\Lambda}_w$.
\paragraph{Approximation} Consider a \ac{BPSK} constallation $\rho_{n,j}\eqdef\frac{\tau_j\invQfunc{\frac{1-\delta}{2}}}{\sqrt{n}}$ for all $j\in\intseq{1}{m}$ with $\{\tau_j\}_{j=1}^m$ defined in a way similar to~\eqref{eq:achv-V-rho-def}. Let $\vct{P}_n=\frac{\invQdelta}{\sqrt{n}}\vct{T}$. For any of the above neighborhoods, the covertness metric at any channel realization $\widetilde{\vct{H}}$ is close to that measured at the corresponding discretization point $\vct{H}$. Precisely, we show that the difference of covertness metric between them vanishes fast with respect to the blocklength $n$ in the following lemma, which is proved in Appendix~\ref{sec:proof-lm-V-compound-approximation}.
\begin{lemma}
\label[lemma]{lemma:V-compound-approximation}
For any $\widetilde{\vct{H}}\in\calS_{J,n}$ and its associated discretization point $\vct{H}\in\calS_n$,
\begin{align}
  \label{eq:V-cmp-code-dist}
    \left|\V{\Qcoden_{\widetilde{\vct{H}}}, \Qinnn}-\V{\Qcoden_{\vct{H}}, \Qinnn}\right|\leq\calO\left(n^{\frac{3}{4}}e^{-n\log 2}\right).
\end{align} 
\end{lemma}
Thus, the covertness metric of any point in $\calS$ can be closely approximated by some discretization point in $\calS_n$ for a sufficiently large $n$.

\paragraph{Existence} 
To show the existence of a compound code applicable for the entire uncertainty set $\calS$, 
note that by the property of supremum and our approximation argument~\eqref{eq:V-cmp-code-dist}, for $n$ sufficiently large, we have
\begin{align}
\label{eq:V-cmp-cvt-cnst-2}
  \left|\sup_{\vct{H}_w\in\calS}\V{\Qcoden_{\vct{H}_w},\Qinnn}-\max_{\vct{H}_w\in\calS_n}\V{\Qcoden_{\vct{H}_w},\Qinnn}\right|
  \leq\calO\left(n^{\frac{3}{4}}e^{-n\log 2}\right).
\end{align}
Accordingly, we choose $\max_{\vct{H}_w\in\calS_n}\V{\Qcoden_{\vct{H}_w},\Qinnn}$ as our optimization constraint by using \eqref{eq:V-cmp-cvt-cnst-2}. This modification only causes a small perturbation in the throughput, which is negligible in the limit of large $n$. Furthermore, we have

\begin{align}
  \label{eq:V-cmp-max-covert-abs}
  &\max_{\vct{H}_w\in\calS_n}\V{\Qcoden_{\vct{H}_w}, \Qinnn}\leq\max_{\vct{H}_w\in\calS_n}\V{\QinfHwn, \Qinnn}+\max_{\vct{H}_w\in\calS_n}\V{\Qcoden_{\vct{H}_w}, \QinfHwn}.
\end{align}
Note that for any $\vct{H},~\widetilde{\vct{H}}\in\calS$ such that $\vct{H}-\widetilde{\vct{H}}\succcurlyeq\vct{0}$ and for $n$ large enough,
    $$\V{\QinfHn, \Qinnn}-\V{\QinftHn, \Qinnn}
    \geq2\Qfunc{\sqrt{\frac{1}{2}\sum_{j=1}^m\frac{\widetilde{\lambda}_j^4\tau_j^2}{4\sigma_w^4}}\invQdelta}-2\Qfunc{\sqrt{\frac{1}{2}\sum_{j=1}^m\frac{\lambda_j^4\tau_j^2}{4\sigma_w^4}}\invQdelta}\geq0,$$ where $\{\lambda_j\}_{j=1}^m$ and $\{\widetilde{\lambda}_j\}_{j=1}^m$ are the diagonal elements of $\vct{\Lambda}$ and $\widetilde{\vct{\Lambda}}$ corresponding to $\vct{H}$ and $\widetilde{\vct{H}}$ defined in ~\eqref{eq:uncertainty-set-def}, respectively. 
To ensure
\begin{align}
  \label{eq:V-cmp-max-cvt-apprx-delta}
  \max_{\vct{H}_w\in\calS_n}\V{Q_{\vct{H}_w}^{\otimes n}, \Qinnn}\leq\delta-\frac{1}{\sqrt{n}},
\end{align}
for large enough $n$, by using~\eqref{eq:V-ch-rsl-cvt-delta} and \eqref{eq:V-ch-rsl-cvt-cnst-1} in Appendix~\ref{sec:proof-lm-achv-V-rsl}, we have the optimization constraint
\begin{align}
\label{eq:V-cmp-cvt-cnst}
    \frac{1}{4\sigma_w^4}\tr{\vct{\Lambda}_0^4\vct{T}^2}\leq 2 - \frac{\bar{C}}{\sqrt{n}\invQdelta}
\end{align}
for some $\bar{C}>0$.
Also, choosing
\begin{equation}
    \label{eq:V-cmp-rsl-code-size}
    \log M_nK_n=\left(1+\xi\right)\frac{1}{2\sigma_w^2}\tr{\vct{\Lambda}_0^2\vct{T}}\sqrt{n}\invQdelta,
\end{equation}
ensures that $\max_{\vct{H}_w\in\calS_n}\expect[\calC]{\V{\Qcoden_{\vct{H}_w}, Q_{\vct{H}_w}^{\otimes n}}}$ vanishes in $\exp\left(-\calO\left(\sqrt{n}\invQdelta\right)\right)$, where it follows from \eqref{eq:V-ch-rsl-code-size} and \eqref{eq:V-ch-rsl-code-dist-exp} in Appendix~\ref{sec:proof-lm-achv-V-rsl}.

We next show that for any \ac{BPSK} random code, $\max_{\vct{H}_w\in\calS_n}\V{\Qcoden_{\vct{H}_w}, \QinfHwn}$ can be upper-bounded by $\max_{\vct{H}_w\in\calS_n}\expect[\calC]{\V{\Qcoden_{\vct{H}_w}, Q_{\vct{H}_w}^{\otimes n}}}$. 
\begin{lemma}
\label[lemma]{lemma:V-cmp-McDiarmid}
For any generated code, set
\begin{align}
    \label{eq:V-cmp-cvt-event}
  \calE\eqdef&\left\{ \max_{\vct{H}_w\in\calS_n}\V{\Qcoden_{\vct{H}_w},\QinfHwn}\leq\max_{\vct{H}_w\in\calS_n}\expect[\calC]{\V{\Qcoden_{\vct{H}_w}, \QinfHwn}}+\alpha_n\right\},
\end{align}
where $\alpha_n=o\left(\frac{1}{\sqrt{n}}\right)$. Then, $\P{\calE}\geq 1-\card{\calS_n}\exp\left(-2M_nK_n\alpha_n^2\right).$
\end{lemma}
\begin{IEEEproof}

We have 
\begin{align}
  &\P{\calE}\labrel\geq{eq:V-cmp-ext-unibd-1}1-\sum_{\vct{H}\in\calS_n}\bbP\Bigg(\V{\Qcoden_{\vct{H}}, \QinfHn}\geq\max_{\vct{H}_w\in\calS_n}\expect[\calC]{\V{\Qcoden_{\vct{H}_w}, \QinfHwn}}+\alpha_n\Bigg)\\
    &\labrel\geq{eq:V-cmp-ext-unibd-2}1-\sum_{\vct{H}\in\calS_n}\P{\V{\Qcoden_{\vct{H}}, \QinfHn}\geq\expect[\calC]{\V{\Qcoden_{\vct{H}}, \QinfHn}}+\alpha_n}\\
    &\labrel\geq{eq:V-cmp-ext-superexp-1}1-\card{\calS_n}\exp\left(-2M_nK_n\alpha_n^2\right),
\end{align}
where \eqref{eq:V-cmp-ext-unibd-1} follows from the union bound, \eqref{eq:V-cmp-ext-unibd-2} follows since $\max_{\vct{H}_w\in\calS_n}\expect[\calC]{\V{\Qcoden_{\vct{H}_w}, \QinfHwn}}\geq\expect[\calC]{\V{\Qcoden_{\vct{H}}, \QinfHn}}$ for any $\vct{H}\in\calS_n$, and \eqref{eq:V-cmp-ext-superexp-1} follows from McDiarmid's Theorem~\cite[Lemma 2]{Tahmasbi2019}. 
\end{IEEEproof}
Hence, we have $\P{\calE}\to 1$ as $n\to\infty$ since, with our choice in \eqref{eq:V-cmp-rsl-code-size},  $\exp\left(-M_nK_n\right)=\exp\left(-\exp\left(\calO\left(\sqrt{n}\invQdelta\right)\right)\right)$ and $\card{\calS_n}=\exp\left(\calO\left(n\right)\right)$. If $n$ is large enough, with overwhelming probability, we can rewrite \eqref{eq:V-cmp-max-covert-abs} as follows:
\begin{align}
  &\max_{\vct{H}_w\in\calS_n}\V{\Qcoden_{\vct{H}_w}, \Qinnn}\leq\max_{\vct{H}_w\in\calS_n}\V{\QinfHwn, \Qinnn}\nonumber\\
    &\quad +\exp\left(-\calO\left(\sqrt{n}\invQdelta\right)\right)+\alpha_n.
\end{align}
As a result, we show that for $n$ large enough, there exists a random code $\calC_c$ generated according to the constraints \eqref{eq:V-cmp-cvt-cnst} and \eqref{eq:V-cmp-rsl-code-size}, which is also a compound covert code for the whole discretized uncertainty set $\calS_n$, i.e.,
\begin{align}
  &\max_{\vct{H}_w\in\calS_n}\V{\Qcoden_{\vct{H}_w}, \Qinnn}\nonumber\\
  &\leq \delta-\frac{1}{\sqrt{n}}+\exp\left(-\calO\left(\sqrt{n}\invQdelta\right)\right)+\alpha_n\\
  &=\delta-\frac{1}{\sqrt{n}}+o\left(\frac{1}{\sqrt{n}}\right)\leq\delta-\frac{D}{\sqrt{n}}
\end{align}
is ensured for sufficiently large $n$ and some $D>0$.

Eventually, combining the above with the triangle inequality and \eqref{eq:V-cmp-cvt-cnst-2}, for $n$ large enough,
we can develop a bound similar to \eqref{eq:V-close-delta} as follows:
\begin{align}
\label{eq:V-cmp-sup-close-delta}
    \sup_{\vct{H}_w\in\calS}\strut\V{\Qcoden_{\vct{H}_w},\Qinnn}\leq\delta-\frac{D}{\sqrt{n}}+\calO\left(n^{\frac{3}{4}}e^{-n\log 2}\right)\leq\delta.
\end{align}
 Therefore, $\calC_c$ is also a compound covert code for the entire uncertainty set $\calS$.

\paragraph{Constellation power design}
The power design follows the same steps as in \eqref{eq:V-opt-1}-\eqref{eq:V-achv-key-msg-tpt}. To find the optimal design point of $\vct{T}$, we first ignore the $\calO\left(\frac{1}{\sqrt{n}\invQdelta}\right)$ term in the \eqref{eq:V-cmp-cvt-cnst} and include it later as a perturbation as in \eqref{eq:V-opt-2-perturbation}. By solving an optimization program similar to \eqref{eq:V-opt-2}, we obtain
\begin{align}
\label{eq:V-cmp-sol-1}
    \limn{\frac{\log M_n}{\sqrt{n}\invQdelta}}=\left(1-\xi\right)\frac{\sigma_w^2}{\sigma_b^2}\sqrt{2\tr{\vct{\Lambda}_b^4\left(\vct{\Lambda}_0^{-1}\right)^4}}.
\end{align}

Therefore, by using \eqref{eq:V-cmp-sup-close-delta}, we can further normalize \eqref{eq:V-cmp-sol-1} and obtain \eqref{eq:compound-opt-tp-stat}. \eqref{eq:compound-opt-k-stat} follows similarly.
\end{IEEEproof}
\section{Comparison and Discussion}
\label{sec:comp-disc}

We now compare our results with the ones obtained when measuring covertness with relative entropy in~\cite{Bendary2019,Bendary2020}. There are several key distinctions between the present work and \cite{Bendary2019,Bendary2020}:
\begin{inparaenum}
\item We analyze covertness in terms of variational distance, which is a more operationally relevant covertness metric, and leads to a higher number of covert bits.
  \item We do not assume that Alice and Bob use a \emph{large amount of key} to create \ac{iid} codewords, and resort instead to a channel resolvability analysis and a conservative amount of secret key $S\in\intseq{1}{K_n}$ shared between Alice and Bob.
    \item We develop a complete characterization of the covert capacity, which is only implicitly defined in~\cite{Bendary2019,Bendary2020} through an optimization problem that depends on the blocklength $n$~\cite[Appendix B]{Bendary2019}.
    \item We do not require the covert bits to be secret. In our opinion, requiring the covert bits to be secret makes the problem closer to a wiretap channel and we want to exclusively focus on covertness.
    \item We do not investigate in depth what happens when channel matrices have non-trivial null spaces, except for a short discussion in Remark~\ref{rmk:full-rank} and Appendix~\ref{sec:furth-expl-full}. In our opinion, these situations are not particularly difficult to analyze because the optimal signaling schemes are rather straightforward.
\end{inparaenum}

Note that \cite{Bendary2019} does not completely fit into our framework since the codebook is assumed to be secret from Willie in~\cite[Theorems 1 and 2]{Bendary2019} and Willie directly observes an \ac{iid} stochastic process. Nevertheless, one can still compare resulting rates.

Note that the result obtained in \cite[Theorem 2]{Bendary2019} is not a closed-form expression, for it involves the characterization of ``normalized KL divergence'' defined in~\cite[(15)]{Bendary2019}. This characterization remains incomplete therein, for the authors do not exactly solve the power allocation problem in~\cite[Appendix B]{Bendary2019}. Hence, the optimal scaling $L$ in~\cite[Theorem 2]{Bendary2019}, which plays the role of the \emph{covert capacity}, has an implicit dependency on the blocklength $n$. To make a fair comparison, we specialize the result in the same scenario as ours (i.e., the signal, the channel matrices, and the \ac{AWGN} are all real number, we consider the same full-rank assumption and our assumption on numbers of antennas, and the covertness requirement is $\avgD{\Qcoden}{\Qinnn}\leq\delta$) to obtain
\begin{subequations}
  \label{eq:D-opt-1}
  \begin{align}
    \label{eq:D-opt-1-obj}
    &\max_{\vct{P}_n\succcurlyeq\vct{0}}\frac{1}{2}\log\det{\vct{I}_m+\frac{1}{\sigma_b^2}\vct{\Lambda}_b\vct{P}_n\vct{\Lambda}_b^\hermittian},\\
    \label{eq:D-opt-1-con}
    &\text{s.t.~} \frac{1}{2}\tr{\frac{1}{\sigma_w^2}\vct{\Lambda}_w\vct{P}_n\vct{\Lambda}_w^\hermittian-\blog{\vct{I}_m+\frac{1}{\sigma_w^2}\vct{\Lambda}_w\vct{P}_n\vct{\Lambda}_w^\hermittian}}\leq\frac{\delta}{n},
  \end{align}
\end{subequations}
By following the same line of reasoning as in \cite[Section V]{Wang2016a} and the sensitivity analysis in the proof of Proposition~\ref{prop:achv-V}, we can actually solve the power allocation problem of~\cite[Appendix B]{Bendary2019} and \eqref{eq:D-opt-1} to obtain the first-order asymptotics. Because the power vanishes with $n$, we also introduce the notation $\rho_{n,j}\eqdef\tau_j\sqrt{\frac{\delta}{n}}$ for all $j\in\intseq{1}{m}$. The power allocation problem, \eqref{eq:D-opt-1-obj}-\eqref{eq:D-opt-1-con}, reduces to
\begin{subequations}
\label{eq:D-opt-2}
    \begin{align}
    \label{eq:D-opt-2-obj}
        &\max_{\vct{T}\succcurlyeq\vct{0}} ~\frac{1}{2\sigma_b^2}\tr{\vct{\Lambda}_b^2\vct{T}}, \\
    \label{eq:D-opt-2-con}
        &\text{s.t.~} \frac{1}{4\sigma_w^4}\tr{\vct{\Lambda}_w^4\vct{T}^2} \leq 1,
    \end{align}
\end{subequations}
where we have ignored higher-order terms vanishing with $n$ because of the sensitivity analysis. The optimal solution to \eqref{eq:D-opt-2} is $\vct{T}=2\sigma_w^2\frac{\vct{\Lambda}_b^2\left(\vct{\Lambda}_w^{-1}\right)^4}{\sqrt{\tr{\vct{\Lambda}_b^4\left(\vct{\Lambda}_w^{-1}\right)^{4}}}}$. We therefore express the covert capacity under a relative entropy metric as follows:
\begin{align}
  \label{eq:D-cvt-cap}
  \limn{\frac{\log M_{n,D}}{\sqrt{n\delta}}}=\frac{\sigma_w^2}{\sigma_b^2}\sqrt{\tr{\vct{\Lambda}_b^4\left(\vct{\Lambda}_w^{-1}\right)^{4}}}.
\end{align}
Hence, the first-order asymptotics of the optimal covert throughput under a relative entropy constraint $\avgD{\Qcoden}{\Qinnn}\leq\delta$ and a variational distance constraint $\V{\Qcoden,\Qinnn}\leq\delta$ can be expressed as 
\begin{align}
  \label{eq:cvt-tpt-delta}
  \limn{\frac{\log M_{n,D}(\delta)}{\sqrt{n}}}&=\frac{\sigma_w^2}{\sigma_b^2}\sqrt{\tr{\vct{\Lambda}_b^4\left(\vct{\Lambda}_w^{-1}\right)^{4}}\delta}\eqdef f_D(\delta),~\text{and}\\
  \limn{\frac{\log M_{n,V}(\delta)}{\sqrt{n}}}&=\frac{\sigma_w^2}{\sigma_b^2}\sqrt{2\tr{\vct{\Lambda}_b^4\left(\vct{\Lambda}_w^{-1}\right)^{4}}}\invQdelta\eqdef f_V(\delta),
\end{align}
respectively. Note that the above results are consistent with the ones for the \ac{AWGN} channel under a relative entropy metric~\cite[Theorem 5]{Wang2016a} and a variational distance metric~\cite{Zhang2019a} if we consider \ac{SISO} channels with unit gain. 

As remarked in~\cite[Remark 2]{Tahmasbi2019}, since $2\V{\Qcoden,\Qinnn}\leq \avgD{\Qcoden}{\Qinnn}$ by Pinsker's inequality, requiring $\avgD{\Qcoden}{\Qinnn}\leq\delta$ is more stringent than requiring $\V{\Qcoden, \Qinnn}\leq\sqrt{\delta/2}$. Consequently,
\begin{align}
  \label{eq:cmp-V-D-opt-cvt-tpt}
  f_D(\delta)\leq f_V(\sqrt{{\delta}/{2}}).
\end{align}
We illustrate the above relation with a simple numerical example. We consider a compound channel case in which both channel matrices are $4\times 4$ real matrices and $\lambda_0=0.05$, the main channel has generalized singular value $\Lambda_b=\diag{0.385, 0.214, 0.172, 0.028}$, and $\sigma_w^2=2\sigma_b^2=0.001$.
\begin{figure}
  \centering
  \includegraphics[scale=0.8]{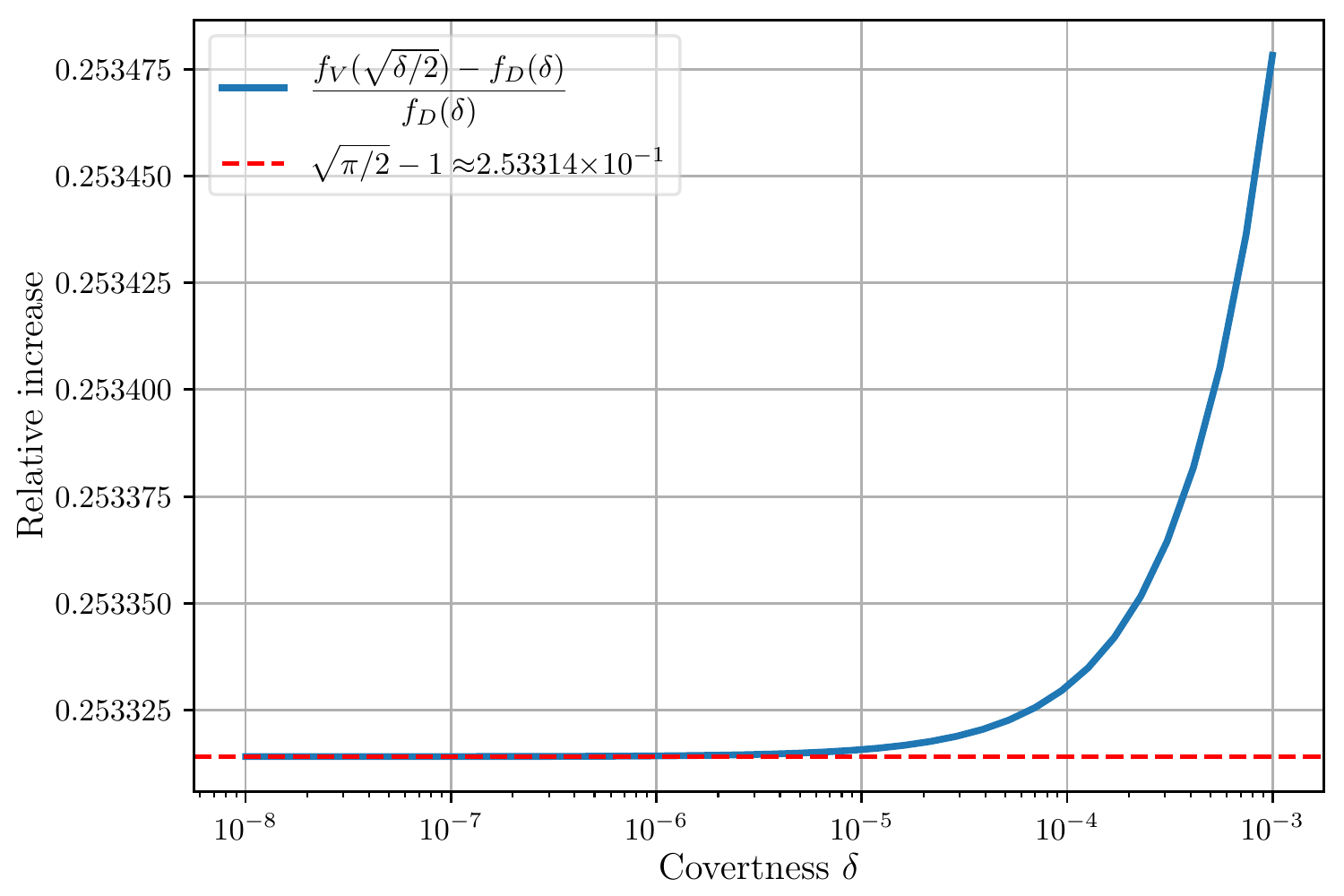}
  \caption{Relative increase of the first-order asymptotics of optimal covert throughput as a function of covertness.}
  \label{fig:cvt-block}
\end{figure}
As shown in Fig.~\ref{fig:cvt-block}, using the variational distance metric results in at least a $25\%$ relative increase in covert throughput. In fact, by the Maclaurin series for the inverse error function $\text{erf}^{-1}$, we have
\begin{align}
  \label{eq:Maclaurin-cvt}
  &\lim_{\delta\to 0}\frac{f_V(\sqrt{{\delta}/{2}})}{f_D(\delta)}=\lim_{\delta\to 0}\frac{\sqrt{2}Q^{-1}(\frac{1-\sqrt{\delta/2}}{2})}{\sqrt{\delta}}=\lim_{\delta\to 0}\frac{2\text{erf}^{-1}(\sqrt{{\delta}/{2}})}{\sqrt{\delta}}\nonumber\\
  &=\lim_{x\to 0}\frac{2\text{erf}^{-1}(x)}{\sqrt{2x^2}}\geq\sqrt{2}\lim_{x\to 0}\frac{\frac{\sqrt{\pi}}{2}x+\calO(x^3)}{x}=\sqrt{\frac{\pi}{2}}.
\end{align} 

\appendices

\section{Further Discussion of the Assumption on Rank and Numbers of Antennas}
\label{sec:furth-expl-full}
In this section, we provide a more detailed discussion of our full-rank assumption for channel matrices $\vct{H}_w$ and $\vct{H}_b$ using a \ac{GSVD} and the analysis of \cite[Section II.A]{Khisti2010}. Without any assumption on the channel matrices, we first define the following subspaces
\begin{align*}
  \calS_b&\eqdef\Null(\vct{H}_b)^{\perp}\cap\Null(\vct{H}_w),\\
  \calS_{b,w}&\eqdef\Null(\vct{H}_b)^{\perp}\cap\Null(\vct{H}_w)^{\perp},\\
  \calS_w&\eqdef\Null(\vct{H}_b)\cap\Null(\vct{H}_w)^\perp,\\
  \calS_n&\eqdef\Null(\vct{H}_b)\cap\Null(\vct{H}_w),
\end{align*}
which correspond to whether the signal in the subspaces can be observed by Bob or Willie. Let
\begin{align*}
  \label{eq:rk}
  m\eqdef\rk{
  \begin{bmatrix}
    \vct{H}_b\\\vct{H}_w
  \end{bmatrix}
},
\end{align*}
 $p\eqdef\dim(\calS_b)$, and $q\eqdef\dim(\calS_{b,w})$. Clearly, $N_a\geq m$, $\dim(\calS_{n})=N_a-m$, and $\dim(\calS_w)=m-p-q$. Both channel matrices can be decomposed with a \ac{GSVD} as  
\begin{equation*}
    \begin{split}
        \vct{H}_b &= \vct{U}_b^\prime\vct{\Sigma}_b[\vct{\Omega}^{-1}\quad\vct{0}_{m\times (N_a-m)}]\vct{\Psi}^\hermittian, \\
        \vct{H}_w &= \vct{U}_w^\prime\vct{\Sigma}_w[\vct{\Omega}^{-1}\quad\vct{0}_{m\times(N_a-m)}]\vct{\Psi}^\hermittian,
    \end{split}
\end{equation*}
where $\vct{\Psi}\in\bbR^{N_a\times N_a}$, $\vct{U}_b^\prime\in\bbR^{N_b\times N_b}$, and $\vct{U}_w^\prime\in\bbR^{N_w\times N_w}$ are orthogonal, $\vct{\Omega}\in\bbR^{m\times m}$ is lower triangular and nonsingular, and
\begin{align*}
  \label{eq:diag-GSVD}
    \vct{\Sigma}_b&=
    \begin{blockarray}{cccc}
      & \scriptstyle{m-p-q} & \scriptstyle{q} & \scriptstyle{p} \\
      \begin{block}{c[ccc]}
        \scriptstyle{N_b-p-q}& \vct{0} & \vct{0} & \vct{0}\\
        \scriptstyle{q} & \vct{0} & \vct{\Lambda}_b & \vct{0}\\
        \scriptstyle{p} & \vct{0} & \vct{0} & \vct{I}\\
      \end{block}
    \end{blockarray}\\
    \vct{\Sigma}_w&=
    \begin{blockarray}{cccc}
      & \scriptstyle{m-p-q} & \scriptstyle{q} & \scriptstyle{p} \\
      \begin{block}{c[ccc]}
        \scriptstyle{m-p-q} & \vct{I} & \vct{0} & \vct{0}\\
        \scriptstyle{q} & \vct{0} & \vct{\Lambda}_w & \vct{0}\\
        \scriptstyle{N_w+p-m} & \vct{0} & \vct{0} & \vct{0}\\
      \end{block}
    \end{blockarray},
\end{align*}
with $\vct{\Lambda}_b=\diag{\{\lambda_{b,j}\}_{j=1}^q}$ and $\vct{\Lambda}_w=\diag{\{\lambda_{w,j}\}_{j=1}^q}$.
Note that the following cases do not contribute to the investigation of the square-root law. 
\begin{inparaenum}
\item If $\calS_b$ is not trivial, Alice can steer her signal in these directions without being detected by Willie, and therefore overcome the square-root law (i.e., the covert capacity as defined in~\eqref{eq:def-pair} is unbounded). The use of these directions contributes nothing to the covertness metric, and a power constraint must be active for the rates to be finite. The optimal power allocation is then the usual water-filling solution. We exclude this case by assuming $p=0$.
\item If $\calS_w$ is not trivial, Alice would avoid using these directions, since the signals in those directions cannot be observed by Bob. We exclude this case by assuming $m=p+q$.
  \item If $\calS_{n}$ is not trivial, Alice would similarly avoid using these directions, and we also exclude this case by assuming $N_a=m$.
\end{inparaenum}
 
In summary, only $\calS_{b,w}$ is relevant to the square-root law. By excluding the above three cases, if $N_a\leq N_b$ and $N_a\leq N_w$, we have the full-rank assumption $m=\rk{\vct{H}_b}=\rk{\vct{H}_w}=N_a$. If $N_a>N_b$ or $N_a>N_w$, some of the mentioned null spaces may not be trivial, and this still falls into the scenarios we point out in Remark~\ref{rmk:full-rank}, which does not affect the investigation of the square-root law. For instance, if $N_a>N_b$, then either $N_a>m$ or $N_a=m$ is true, and the former case corresponds to a non-trivial $\calS_n$. If we also have $N_a=m$, since $m=N_a>N_b\geq q$, at least one of $m>p+q$ or $p>0$ is true, which corresponds to the non-trivial $\calS_w$ or $\calS_b$. Hence, we also impose the assumption that both Bob and Willie possess more antennas than Alice. The assumptions on numbers of antennas and rank are without loss of generality, and the reason is simply to exclude some perhaps less interesting cases to reveal the constant before the square-root instead of preventing any technical difficulty.
\color{black}
\section{Proof of \Cref{lemma:code-min-power-bound}}
\label{sec:pf-lm-code-min-power-bound}
\begin{IEEEproof}
Under the null hypothesis $H_0$, for all $i\in\intseq{1}{n}$ $\vct{Z}_i\sim\N{\vct{0}}{\sigma_w^2\vct{I}_{N_w}}$, so that $\hat{\vct{Z}}_i\sim\N{\vct{0}}{\sigma_w^2\vct{\Lambda}_b^2\left(\vct{\Lambda}_w^{-1}\right)^2}\in\bbR^m$. 
Hence, we have the following statistics:
\begin{align}
    \mu_{0}&=\sum_{i=1}^n\expect[\Qinn]{S_i}=n\sigma_w^2\tr{\vct{\Lambda}_b^2\left(\vct{\Lambda}_w^{-1}\right)^2},\\
    \sigma_0^2&=\sum_{i=1}^n\Var{S_i}=2n\sigma_w^4\tr{\vct{\Lambda}_b^4\left(\vct{\Lambda}_w^{-1}\right)^4},\\
    t_{0,i}&=\expect[Q_0]{|S_i-\mu_{0,i}|^3}=\calO\left(1\right),~t_0=\sum_{i=1}^nt_{0,i}=\calO\left(n\right),
\end{align}
where $\mu_{0, i}\eqdef\expect[\Qinn]{S_i}.$
We use the Berry-Esseen Theorem to obtain an upper bound for the probability of false alarm as follows:
\begin{align}
  \alpha&=\P[H_0]{\sum_{i=1}^nS_i\geq \tau}\\
  &\leq\Qfunc{\frac{\tau-n\sigma_w^2\tr{\vct{\Lambda}_b^2\left(\vct{\Lambda}_w^{-1}\right)^2}}{\sqrt{2n\tr{\vct{\Lambda}_b^4\left(\vct{\Lambda}_w^{-1}\right)^4}}\sigma_w^2}}+\frac{6t_0}{\sigma_0^3}\\
    &\leq\Qfunc{\frac{\tau-n\sigma_w^2\tr{\vct{\Lambda}_b^2\left(\vct{\Lambda}_w^{-1}\right)^2}}{\sqrt{2n\tr{\vct{\Lambda}_b^4\left(\vct{\Lambda}_w^{-1}\right)^4}}\sigma_w^2}}+\frac{B_0}{\sqrt{n}}.
\end{align}
The bound on the probability of false alarm~\eqref{eq:V-conv-alpha} follows by applying the threshold $\tau=\frac{P_*}{2}+n\sigma_w^2\tr{\vct{\Lambda}_b^2\left(\vct{\Lambda}_w^{-1}\right)^2}.$

Similarly, under the hypothesis $H_1$, we know that given a codeword $\vct{x}^{\left(k\right)n}$ transmitted over the channel $\Wwn$, for every $i\in\intseq{1}{n}$ $\vct{Z}_i|\vct{X}_i=\vct{x}_i^{(k)}\sim\N{\vct{H}_w\vct{x}_i^{\left(k\right)}}{\sigma_w^2\vct{I}_{N_w}}$, so that $\hat{\vct{Z}}_i|\parVarX_i=\parVarx_i^{(k)}\sim\N{\vct{\Lambda}_b\parVarx_i^{(k)}}{\sigma_w^2\vct{\Lambda}_b^2\left(\vct{\Lambda}_w^{-1}\right)^2}$, where $\parVarx_i^{(k)}=\vct{V}^\hermittian\vct{x}_i^{(k)}$. 
Let $\parVarx^{(k)n}\eqdef\vct{V}^\hermittian\vct{x}^{(k)n}$ and $\vct{P}^{(k)}\eqdef\sum_{i=1}^n\parVarx_i^{(k)}\parVarx_i^{(k)\hermittian}$. Hence, we have the following statistics:

\begin{align}
  \label{eq:V-conv-mu-1}
  \mu_{1}^{\left(k\right)}&=\sum_{i=1}^n\expect[\widehat{Q}]{S_i|\vct{X}_i
                            =\vct{x}_i^{\left(k\right)}}=\tr{\vct{\Lambda}_b^2\vct{P}^{(k)}}+n\sigma_w^2\tr{\vct{\Lambda}_b^2\left(\vct{\Lambda}_w^{-1}\right)^2},\\
  \label{eq:V-conv-sigma-1}
  \sigma_{1}^{\left(k\right)2} &= \sum_{i=1}^n\Var{S_i|\vct{X}_i
                                 =\vct{x}_i^{\left(k\right)}}= 4\sigma_w^2\tr{\vct{\Lambda}_b^4\left(\vct{\Lambda}_w^{-1}\right)^2\vct{P}^{\left(k\right)}}+2n\sigma_w^4\tr{\vct{\Lambda}_b^4\left(\vct{\Lambda}_w^{-1}\right)^4},\\
  t_{1,i}^{\left(k\right)}&=\expect[\widehat{Q}]{|S_i-\mu_{1,i}^{(k)}|^3|\vct{X}_i=\vct{x}_i^{\left(k\right)}}=\calO\left(1\right),\nonumber\\
  t_{1}^{\left(k\right)}&=\sum_{i=1}^n t_{1,i}^{\left(k\right)}=\calO\left(n\right),
\end{align}
where $\mu_{1, i}^{(k)}\eqdef\expect[\widehat{Q}]{S_i|\vct{X}_i=\vct{x}_i^{\left(k\right)}}.$ We provide the algebraic details of \eqref{eq:V-conv-mu-1} and \eqref{eq:V-conv-sigma-1} in Appendix~\ref{sec:algebraic-details}.
We use again the Berry-Esseen Theorem to obtain an upper bound for the probability of missed detection. For the $k$-th codeword, \textcolor{black}{by defining $\beta^{\left(k\right)} = \P[H_1]{\sum_{i=1}^nS_i<\tau|\vct{X}^n=\vct{x}^{\left(k\right)n}}$}, we have \eqref{eq:V-conv-beta-k-init},
\begin{figure*}[!t]
  \begin{align}
    \beta^{\left(k\right)}&\leq \Qfunc{\frac{\tr{\vct{\Lambda}_b^2\vct{P}^{(k)}}+n\sigma_w^2\tr{\vct{\Lambda}_b^2\left(\vct{\Lambda}_w^{-1}\right)^2}-\tau}{\sqrt{4\sigma_w^2\tr{\vct{\Lambda}_b^4\left(\vct{\Lambda}_w^{-1}\right)^2\vct{P}^{\left(k\right)}}+2n\sigma_w^4\tr{\vct{\Lambda}_b^4\left(\vct{\Lambda}_w^{-1}\right)^4}}}}+\frac{6t_1^{\left(k\right)}}{\sigma_1^{\left(k\right)3}}\\
    \label{eq:V-conv-beta-k-init}
    &\labrel\leq{eq:MD-trace-ineq} \Qfunc{\frac{\tr{\vct{\Lambda}_b^2\vct{P}^{(k)}}+n\sigma_w^2\tr{\vct{\Lambda}_b^2\left(\vct{\Lambda}_w^{-1}\right)^2}-\tau}{\sqrt{4\sigma_w^2\tr{\vct{\Lambda}_b^2\vct{P}^{(k)}}\tr{\vct{\Lambda}_b^2\left(\vct{\Lambda}_w^{-1}\right)^2}+2n\sigma_w^4\tr{\vct{\Lambda}_b^4\left(\vct{\Lambda}_w^{-1}\right)^4}}}}+\frac{B_1}{\sqrt{n}},
\end{align}
\hrulefill
\vspace*{4pt}
\end{figure*}
where \eqref{eq:MD-trace-ineq} follows \textcolor{black}{since $\vct{\Lambda}_b$, $\vct{\Lambda}_w$, $\vct{P}^{k}\succeq\vct{0}$,} and since $\tr{\vct{\Lambda}_b^4\left(\vct{\Lambda}_w^{-1}\right)^2\vct{P}^{(k)}}\leq\tr{\vct{\Lambda}_b^2\left(\vct{\Lambda}_w^{-1}\right)^2}\tr{\vct{\Lambda}_b^2\vct{P}^{(k)}}$. By setting the detection threshold to $\tau=\frac{P_*}{2}+n\sigma_w^2\tr{\vct{\Lambda}_b^2\left(\vct{\Lambda}_w^{-1}\right)^2}$, we further obtain \eqref{eq:MD-apprx-k},
\begin{figure*}[!t]
\begin{align}
    \beta^{\left(k\right)} &\labrel\leq{eq:MD-Pmin} \Qfunc{\frac{\frac{P_*}{2}}{\sqrt{4P_*\sigma_w^2\tr{\vct{\Lambda}_b^2\left(\vct{\Lambda}_w^{-1}\right)^2}+2n\sigma_w^4\tr{\vct{\Lambda}_b^4\left(\vct{\Lambda}_w^{-1}\right)^4}}}}+\frac{B_1}{\sqrt{n}} \\
    &=\Qfunc{\frac{P_*}{2\sqrt{2n\tr{\vct{\Lambda}_b^4\left(\vct{\Lambda}_w^{-1}\right)^4}}\sigma_w^2}\frac{1}{\sqrt{1+\frac{2P_*\tr{\vct{\Lambda}_b^2\left(\vct{\Lambda}_w^{-1}\right)^2}}{n\sigma_w^2\tr{\vct{\Lambda}_b^4\left(\vct{\Lambda}_w^{-1}\right)^4}}}}} + \frac{B_1}{\sqrt{n}} \\
    &\labrel\leq{eq:MD-sqrt-apprx} \Qfunc{\frac{P_*}{2\sqrt{2n\tr{\vct{\Lambda}_b^4\left(\vct{\Lambda}_w^{-1}\right)^4}}\sigma_w^2}\left(1-\frac{P_*\tr{\vct{\Lambda}_b^2\left(\vct{\Lambda}_w^{-1}\right)^2}}{n\sigma_w^2\tr{\vct{\Lambda}_b^4\left(\vct{\Lambda}_w^{-1}\right)^4}}\right)} + \frac{B_1}{\sqrt{n}} \\
    \label{eq:MD-apprx-k}
    &\labrel\leq{eq:MD-Q-apprx1} \Qfunc{\frac{P_*}{2\sqrt{2n\tr{\vct{\Lambda}_b^4\left(\vct{\Lambda}_w^{-1}\right)^4}}\sigma_w^2}} + \frac{P_*^2\tr{\vct{\Lambda}_b^2\left(\vct{\Lambda}_w^{-1}\right)^2}}{4\sqrt{\pi}n^{3/2}\tr{\vct{\Lambda}_b^4\left(\vct{\Lambda}_w^{-1}\right)^4}^{3/2}\sigma_w^4} + \frac{B_1}{\sqrt{n}},
\end{align}
\hrulefill
\vspace*{4pt}
\end{figure*}
where \eqref{eq:MD-Pmin} follows since $\tr{\vct{\Lambda}_b^2\vct{P}^{(k)}}-\frac{P_*}{2}\geq\frac{\tr{\vct{\Lambda}_b^2\vct{P}^{(k)}}}{2}\geq\frac{P_*}{2}$ and $\frac{\tr{\vct{\Lambda}_b^2\vct{P}^{(k)}}}{\sqrt{a\tr{\vct{\Lambda}_b^2\vct{P}^{(k)}}+b}}\geq\frac{P_*}{\sqrt{aP_*+b}},$ $\forall a,b>0$, \eqref{eq:MD-sqrt-apprx} follows from $\frac{1}{\sqrt{1+x}}\geq1-\frac{x}{2},~\forall x>0$, and \eqref{eq:MD-Q-apprx1} follows from $\Qfunc{x-y}=\Qfunc{x}+\int_{x-y}^x\frac{1}{\sqrt{2\pi}}\exp{\left(-\frac{u^2}{2}\right)}du\leq\Qfunc{x}+\frac{y}{\sqrt{2\pi}}~\forall~0<y<x$. Note that the upper bound \eqref{eq:MD-apprx-k} is independent of the codeword $\vct{x}^{\left(k\right)n}$, and hence we can use the bound on all the $\beta^{\left(k\right)}$'s. Therefore, we obtain
\begin{align}
    \beta &= \P[H_1]{\sum_{i=1}^nS_i<\tau} = \frac{1}{\card{\calM}}\sum_{\vct{x}^{\left(k\right)n}\in\calM}\beta^{\left(k\right)}
            \nonumber\\
          &\leq\Qfunc{\frac{P_*}{2\sqrt{2n\tr{\vct{\Lambda}_b^4\left(\vct{\Lambda}_w^{-1}\right)^4}}\sigma_w^2}} \nonumber\\
  &\quad+ \frac{P_*^2\tr{\vct{\Lambda}_b^2\left(\vct{\Lambda}_w^{-1}\right)^2}}{4\sqrt{\pi}n^{3/2}\tr{\vct{\Lambda}_b^4\left(\vct{\Lambda}_w^{-1}\right)^4}^{3/2}\sigma_w^4} + \frac{B_1}{\sqrt{n}}.
\end{align}
\end{IEEEproof}

\section{Algebraic Details of Equations \eqref{eq:V-conv-mu-1} and \eqref{eq:V-conv-sigma-1}}
\label{sec:algebraic-details}
In this appendix, since we only consider a specific codeword $\vct{x}^{(k)n}$, we suppress the codeword index $k$ to simplify the notation. To make the calculation more precise, we use $\hat{z}_{ij}$ and $\tilde{x}_{ij}$ to represent the component $j\in\intseq{1}{m}$ of vector $\hat{\vct{z}}_i$ and $\tilde{\vct{x}}_i$, respectively. Since $\hat{\vct{Z}}_i|\parVarX_i=\parVarx_i\sim\N{\vct{\Lambda}_b\parVarx_i}{\sigma_w^2\vct{\Lambda}_b^2\left(\vct{\Lambda}_w^{-1}\right)^2}$, we have $\hat{Z}_{ij}|\tilde{X}_{ij}=\tilde{x}_{ij}\sim\N{\lambda_{b,j}\tilde{x}_{ij}}{\sigma_w^2\lambda_{b,j}^2\lambda_{w,j}^{-2}}$. Then, 
\begin{align}
  \mu_{1i}&\eqdef\expect[\widehat{Q}]{S_i\vert\vct{X}_i=\vct{x}_i}=\expect[\widehat{Q}]{S_i\vert\tilde{\vct{X}}_i=\tilde{\vct{x}}_i}\\
          &=\sum_{j=1}^m\expect[\Qcode]{\hat{Z}^2_{ij}\vert \tilde{X}_{ij}=\tilde{x}_{ij}}\\
  &= \sum_{j=1}^m\left(\lambda_{b,j}^2\tilde{x}_{ij}^2+\sigma_w^2\lambda_{b,j}^2\lambda_{w,j}^{-2}\right),
\end{align}
and we therefore have
\begin{align}
  \mu_1&=\sum_{i=1}^n\mu_{1i}\\
       &=\sum_{j=1}^m\lambda_{b,j}^2\left(\sum_{i=1}^n\tilde{x}_{ij}^2\right)+n\sigma^2_w\sum_{j=1}^m\lambda_{b,j}^2\lambda_{w,j}^{-2}\\
  &= \tr{\vct{\Lambda}_b^2\vct{P}}+n\sigma_w^2\tr{\vct{\Lambda}_b^2\left(\vct{\Lambda}_w^{-1}\right)^2}.
\end{align}
Note that $\vct{P}\eqdef\sum_{i=1}^n\tilde{\vct{x}}_i\tilde{\vct{x}}_i^{\hermittian}$ has $\sum_{i=1}^n\tilde{x}_{ij}^2$'s as its diagonal elements.

Next, we turn to the analysis of the variance $\sigma_1^2$. First, we have
\begin{align}
&\expect[\widehat{Q}]{S_i^2\vert\vct{X}_i=\vct{x}_i}=\expect[\widehat{Q}]{S_i^2\vert\tilde{\vct{X}}_i=\tilde{\vct{x}}_i}\\
  &=\expect[\Qcode]{\sum_{j=1}^m\sum_{r=1}^m\hat{Z}^2_{ij}\hat{Z}^2_{ir}\vert \tilde{X}_{ij}=\tilde{x}_{ij},\tilde{X}_{ir}=\tilde{x}_{ir}}\\
  &=\sum_{j=r}\expect[\Qcode]{\hat{Z}^4_{ij}\vert \tilde{X}_{ij}=\tilde{x}_{ij}}+\sum_{j\neq r}\expect[\Qcode]{\hat{Z}^2_{ij}\hat{Z}^2_{ir}\vert \tilde{X}_{ij}=\tilde{x}_{ij},\tilde{X}_{ir}=\tilde{x}_{ir}}\\
  &=\sum_{j=1}^m\left(\lambda_{b,j}^4\tilde{x}_{ij}^4+6\sigma_w^2\lambda_{b,j}^4\lambda_{w,j}^{-2}\tilde{x}_{ij}^2+3\lambda_{b,j}^4\lambda_{w,j}^{-4}\sigma_w^4\right)+\sum_{j\neq r}\lambda_{b,j}^2\lambda_{w,j}^{-2}\lambda_{b,r}^2\lambda_{w,r}^{-2}\sigma_w^4.
\end{align}
On the other hand,
\begin{align}
  &\expect[\widehat{Q}]{S_i\vert\vct{X}_i=\vct{x}_i}^2=\left(\sum_{j=1}^m\expect[\Qcode]{\hat{Z}^2_{ij}\vert \tilde{X}_{ij}=\tilde{x}_{ij}}\right)^2\\
  &=\sum_{j=r}\expect[\Qcode]{\hat{Z}^2_{ij}\vert \tilde{X}_{ij}=\tilde{x}_{ij}}^2+\sum_{j\neq r}\expect[\Qcode]{\hat{Z}^2_{ij}\vert \tilde{X}_{ij}=\tilde{x}_{ij}}\expect[\Qcode]{\hat{Z}^2_{ir}\vert \tilde{X}_{ir}=\tilde{x}_{ir}}\\
  &=\sum_{j=1}^m\left(\lambda_{b,j}^4\tilde{x}_{ij}^4+2\sigma_w^2\lambda_{b,j}^4\lambda_{w,j}^{-2}\tilde{x}_{ij}^2+\lambda_{b,j}^4\lambda_{w,j}^{-4}\sigma_w^4\right)+\sum_{j\neq r}\lambda_{b,j}^2\lambda_{w,j}^{-2}\lambda_{b,r}^2\lambda_{w,r}^{-2}\sigma_w^4.
\end{align}
Therefore, we have
\begin{align}
  \sigma_{1i}^2&\eqdef\expect[\widehat{Q}]{S_i^2\vert\vct{X}_i=\vct{x}_i}-\expect[\widehat{Q}]{S_i\vert\vct{X}_i=\vct{x}_i}^2\\
               &=4\sigma_w^2\sum_{j=1}^m\lambda_{b,j}^4\lambda_{w,j}^{-2}\tilde{x}_{ij}^2+2\sigma_w^4\sum_{j=1}^m\lambda_{b,j}^4\lambda_{w,j}^{-4},
\end{align}
so that $\sigma_1^2=4\sigma_w^2\tr{\vct{\Lambda}_b^4\left(\vct{\Lambda}_w^{-1}\right)^2\vct{P}}+2n\sigma_w^4\tr{\vct{\Lambda}_b^4\left(\vct{\Lambda}_w^{-1}\right)^4}$.

\section{Proof of \Cref{lemma:V-conv-sub-code-power}}
\label{sec:proof-lm-V-conv-sub-code-power}

\begin{IEEEproof}
We partition the code $\calC$ into two different sub-codes, a low-power sub-code $\calC^{\left(\ell\right)}$ and a high-power sub-code $\calC^{\left(h\right)}$, where 
$    \calC^{\left(\ell\right)}\eqdef\{\vct{x}^n\in\calC:\norm[F]{\vct{H}_b\vct{x}^n}^2\leq A\sqrt{n}\},~\calC^{\left(h\right)}\eqdef\calC\backslash\calC^{\left(\ell\right)}.$
The output distributions induced by these two sub-codes are
\begin{align}
  \widehat{Q}^{\left(\ell\right)}\left(\vct{z}^n\right)=\frac{1}{\card{\calC^{\left(\ell\right)}}}\sum_{\vct{x}^n\in\calC^{\left(\ell\right)}}\Wwn\left(\vct{z}^n|\vct{x}^n\right),\nonumber\\
  ~\text{and}~\widehat{Q}^{\left(h\right)}\left(\vct{z}^n\right)=\frac{1}{\card{\calC^{\left(h\right)}}}\sum_{\vct{x}^n\in\calC^{\left(h\right)}}\Wwn\left(\vct{z}^n|\vct{x}^n\right),
\end{align}
respectively. Note that $\Qcoden=\frac{\card{\calC^{\left(\ell\right)}}}{\card{\calC}}\widehat{Q}^{\left(\ell\right)}+\frac{\card{\calC^{\left(h\right)}}}{\card{\calC}}\widehat{Q}^{\left(h\right)}$. For a code $\calC$ such that $\V{\Qcoden,\Qinnn}\leq\delta$, we have
\begin{align}
    \delta &\geq\V{\Qcoden,\Qinnn}\\
    &\labrel\geq{eq:sub-code-tri-ineq} \frac{\card{\calC^{\left(h\right)}}}{\card{\calC}}\V{\widehat{Q}^{\left(h\right)},\Qinnn}-\frac{\card{\calC^{\left(\ell\right)}}}{\card{\calC}}\V{\widehat{Q}^{\left(\ell\right)},\Qinnn}\\
    &= \V{\widehat{Q}^{\left(h\right)},\Qinnn} -\frac{\card{\calC^{\left(\ell\right)}}}{\card{\calC}}\left(\V{\widehat{Q}^{\left(h\right)},\Qinnn}+\V{\widehat{Q}^{\left(\ell\right)},\Qinnn}\right)\\
    & \labrel\geq{eq:sub-code-V-less-1}\V{\widehat{Q}^{\left(h\right)},\Qinnn} -2\frac{\card{\calC^{\left(\ell\right)}}}{\card{\calC}}\\
             &\labrel\geq{eq:sub-code-plug-in-A} 1 - 2\Qfunc{\frac{A\sqrt{n}}{2\sqrt{2n\tr{\vct{\Lambda}_b^4\left(\vct{\Lambda}_w^{-1}\right)^4}}\sigma_w^2}}-\frac{A^2n\tr{\vct{\Lambda}_b^2\left(\vct{\Lambda}_w^{-1}\right)^2}}{4\sqrt{\pi}n^{3/2}\tr{\vct{\Lambda}_b^4\left(\vct{\Lambda}_w^{-1}\right)^4}^{3/2}\sigma_w^4}-\frac{B_0+B_1}{\sqrt{n}}-2\frac{\card{\calC^{\left(\ell\right)}}}{\card{\calC}} \\
             &= \delta+\frac{2\nu^2\tr{\vct{\Lambda}_b^2\left(\vct{\Lambda}_w^{-1}\right)^2}}{4\sqrt{\pi n}\tr{\vct{\Lambda}_b^4\left(\vct{\Lambda}_w^{-1}\right)^4}^{3/2}\sigma_w^4}-\frac{B_0+B_1}{\sqrt{n}}-\frac{A^2\tr{\vct{\Lambda}_b^2\left(\vct{\Lambda}_w^{-1}\right)^2}}{4\sqrt{\pi n}\tr{\vct{\Lambda}_b^4\left(\vct{\Lambda}_w^{-1}\right)^4}^{3/2}\sigma_w^4}+2\gamma_n-2\frac{\card{\calC^{\left(\ell\right)}}}{\card{\calC}}\\
    &\labrel\geq{eq:sub-code-specify-nu} \delta+2\gamma_n-2\frac{\card{\calC^{\left(\ell\right)}}}{\card{\calC}}
\end{align}
where \eqref{eq:sub-code-tri-ineq} follows from 
\begin{align}
  &\frac{\card{\calC^{\left(h\right)}}}{\card{\calC}}\V{\widehat{Q}^{\left(h\right)},\Qinnn}\\
  &=\frac{1}{2}\norm[1]{\left(\Qcoden-\Qinnn\right)-\frac{\card{\calC^{\left(\ell\right)}}}{\card{\calC}}\left(\widehat{Q}^{\left(\ell\right)}-\Qinnn\right)} \\
&\leq\V{\Qcoden,\Qinnn}+\frac{\card{\calC^{\left(\ell\right)}}}{\card{\calC}}\V{\widehat{Q}^{\left(\ell\right)},\Qinnn},
\end{align}
\eqref{eq:sub-code-V-less-1} follows since the variational distance between any two distributions is upper bounded by $1$, \eqref{eq:sub-code-plug-in-A} follows from~\eqref{eq:V-converse-apprx}, and \eqref{eq:sub-code-specify-nu} follows by choosing $\nu$ to satisfy
\begin{align}
  \frac{2\nu^2\tr{\vct{\Lambda}_b^2\left(\vct{\Lambda}_w^{-1}\right)^2}}{4\sqrt{\pi n}\tr{\vct{\Lambda}_b^4\left(\vct{\Lambda}_w^{-1}\right)^4}^{3/2}\sigma_w^4}-\frac{B_0+B_1}{\sqrt{n}}
  -\frac{A^2\tr{\vct{\Lambda}_b^2\left(\vct{\Lambda}_w^{-1}\right)^2}}{4\sqrt{\pi n}\tr{\vct{\Lambda}_b^4\left(\vct{\Lambda}_w^{-1}\right)^4}^{3/2}\sigma_w^4} > 0.
\end{align}
Hence, we can bound the cardinality of the low-power sub-code $\calC^{\left(\ell\right)}$ from below as 
    $\card{\calC^{\left(\ell\right)}}\geq\gamma_n\card{\calC},$
which shows the existence of such a low-power sub-code.
\end{IEEEproof}

\section{Proof of \Cref{lemma:achv-V-rel}}
\label{sec:proof-lm-achv-V-rel}

\begin{IEEEproof}
We specialize~\cite[Lemma 3]{Bloch2016} and \cite[Lemma 1]{Tahmasbi2019} into the following lemma.
\begin{lemma}
  \label[lemma]{lemma:ch-rel}
For any $\gamma>0$,
\begin{align}
\label{eq:V-ch-reliability-1}
    \expect{\bar{P}^{(n)}_{e}}\leq M_{n}e^{-\gamma}\left(1+\expect[\Pinfn]{\frac{\Pinfn\left(\vct{Y}^n\right)}{\Pinnn\left(\vct{Y}^n\right)}}\right)+\P[\PiQnn\Wbn]{\log\frac{\Wbn\left(\vct{Y}^n|\vct{X}^n\right)}{\Pinnn\left(\vct{Y}^n\right)}\leq\gamma},
\end{align}
where $\bar{P}^{(n)}_{e}$ is the average probability of error.
\end{lemma}
We first analyze the first term on the right-hand side of \eqref{eq:V-ch-reliability-1} as follows:
\begin{align} &\expect[\Pinf]{\frac{\Pinf\left(\vct{Y}\right)}{\Pinn\left(\vct{Y}\right)}}\labrel={eq:V-ch-rel-sub-ch}\prod_{j=1}^m\expect[\parjPinf]{\frac{\parjPinf\left(\tilde{Y}_j\right)}{\parjPinn\left(\tilde{Y}_j\right)}}
=\prod_{j=1}^m\cosh\left(\frac{\lambda_{b,j}^2\rho_{n,j}}{\sigma_b^2}\right) \labrel\leq{eq:V-ch-rel-cosh-ineq}\exp\left(\frac{\sum_{j=1}^m\lambda_{b,j}^4\rho_{n,j}^2}{2\sigma_b^4}\right),
\end{align}
where \eqref{eq:V-ch-rel-sub-ch} follows from the facts that we apply the orthogonal transform $\vct{U}_b^{\prime\hermittian}$ to the observation $\vct{Y}$, each sub-channel is independent, and this mapping is one-to-one and onto. Note that after the orthogonal transform, we could truncate the last $N_b-m$ components, since they contain pure noise (as they correspond to the null space of $\vct{H}_b^\hermittian$) and thus do not affect the decoding process. \eqref{eq:V-ch-rel-cosh-ineq} follows from $\cosh(x)\leq e^{\frac{x^2}{2}}$ and the exponential property.
Therefore, we know that 
\begin{equation}
\label{eq:V-ch-rel-penalty}
\expect[\Pinfn]{\frac{\Pinfn\left(\vct{Y}^n\right)}{\Pinnn\left(\vct{Y}^n\right)}}\leq\exp\left(\frac{n\sum_{j=1}^m\lambda_{b,j}^4\rho_{n,j}^2}{2\sigma_b^4}\right)=\calO\left(1\right).
\end{equation}
Next, we turn to analyze the last term of \eqref{eq:V-ch-reliability-1}. Similarly, with the above orthogonal transform, note that
\begin{align}
    &\log\frac{\Wbn\left(\vct{Y}^n|\vct{X}^n\right)}{\Pinnn\left(\vct{Y}^n\right)}=\sum_{j=1}^m\log\frac{\parjWbn\left(\tilde{Y}^n_j|\tilde{X}^n_j\right)}{\parjPinnn\left(\tilde{Y}^n_j\right)}\\
    &=\sum_{i=1}^n\sum_{j=1}^m\left(\frac{\lambda_{b,j}\tilde{X}_{ij}\tilde{Y}_{ij}}{\sigma_b^2}-\frac{\lambda_{b,j}^2\tilde{X}_{ij}^2}{2\sigma_b^2}\right).
\end{align}
Since each sub-channel is independent and $\tilde{Y}_{ij}|\tilde{X}_{ij}=\tilde{x}_{ij}\sim\N{\lambda_{b,j}\tilde{x}_{ij}}{\sigma_b^2}$ and $\tilde{x}_{ij}\in\{-a_{n,j},a_{n,j}\}$ for every $j$, we have 
  $\sum_{j=1}^m\log\frac{\parjWb\left(\tilde{Y}_{ij}|\tilde{X}_{ij}=\tilde{x}_{ij}\right)}{\parjPinn\left(\tilde{Y}_{ij}\right)}\sim\sum_{j=1}^m\N{\frac{\lambda_{b,j}^2\rho_{n,j}}{2\sigma_b^2}}{\frac{\lambda_{b,j}^2\rho_{n,j}}{\sigma_b^2}}$.
Therefore, by setting 
    $\gamma=\left(1-\epsilon\right)n\sum_{j=1}^m\frac{\lambda_{b,j}^2\rho_{n,j}}{2\sigma_b^2},$
where $\epsilon\in(0,1)$, and using Hoeffding's inequality, we have
\begin{align}
  &\bbP_{W_{\parVarY|\parVarX=\tilde{\vct{x}}^n}^{\otimes n}}\left(\sum_{i=1}^n\sum_{j=1}^m\log\frac{\parjWb\left(\tilde{Y}_{ij}|\tilde{x}_{ij}\right)}{\parPinn\left(\tilde{Y}_{ij}\right)} \leq n\left(1-\epsilon\right)\sum_{j=1}^m\frac{\lambda_{b,j}^2\rho_{n,j}}{2\sigma_b^2}\right)\leq\exp{\left(-n\sum_{j=1}^m\frac{\epsilon^2\lambda_{b,j}^2\rho_{n,j}}{8\sigma_b^2}\right)}.
\end{align}
Then, we have
\begin{align} 
  &\quad\P[\PiQnn\Wbn]{\log\frac{\Wbn\left(\vct{Y}^n|\vct{X}^n\right)}{\Pinnn\left(\vct{Y}^n\right)}\leq\gamma}\\
               &=\sum_{\tilde{\vct{x}}^n\in\prod_{j=1}^m\left\{-a_{n,j},a_{n,j}\right\}^{ n}}\PiPnn\left(\tilde{\vct{x}}^n\right)\bbP_{W_{\parVarY|\parVarX=\tilde{\vct{x}}^n}^{\otimes n}}\left(\sum_{i=1}^n\sum_{j=1}^m\log\frac{\parjWb\left(\tilde{Y}_{ij}|\tilde{x}_{ij}\right)}{\parPinn\left(\tilde{Y}_{ij}\right)}\leq n\left(1-\epsilon\right)\sum_{j=1}^m\frac{\lambda_{b,j}^2\rho_{n,j}}{2\sigma_b^2}\right)\\
  \label{eq:V-ch-rel-error-prob-1}&\leq\exp{\left(-n\sum_{j=1}^m\frac{\epsilon^2\lambda_{b,j}^2\rho_{n,j}}{8\sigma_b^2}\right)}.
\end{align}
Eventually, by combining~\eqref{eq:V-ch-reliability-1}, \eqref{eq:V-ch-rel-penalty}, and \eqref{eq:V-ch-rel-error-prob-1}, we have
\begin{equation}
    \label{eq:V-ch-rel-error-prob-2}
    \expect{\bar{P}^{(n)}_{e}}\leq\exp{\left(-n\sum_{j=1}^m\frac{\epsilon^2\lambda_{b,j}^2\rho_{n,j}}{8\sigma_b^2}\right)}+M_{n}e^{-\gamma}(1+\calO(1)).
\end{equation}
Hence, by using \eqref{eq:achv-V-rho-def}, if we choose
\begin{align}
    \label{eq:V-ch-rel-msg-size}
    \log M_{n}&=\left(1-\omega\right)\left(1-\epsilon\right)n\sum_{j=1}^m\frac{\lambda_{b,j}^2\rho_{n,j}}{2\sigma_b^2}\nonumber\\
    &=\left(1-\xi\right)\sum_{j=1}^m\frac{\lambda_{b,j}^2\tau_j\sqrt{n}\invQdelta}{2\sigma_b^2},
\end{align}
where $\omega\in (0, 1)$ and $\xi=\frac{1}{2}[(1+\omega)(1+\epsilon)-(1-\omega)(1-\epsilon)]>0$, the result follows.
\end{IEEEproof}

\section{Proof of \Cref{lemma:achv-V-rsl}}
\label{sec:proof-lm-achv-V-rsl}

\begin{IEEEproof}
In the following, we apply the triangle inequality to upper-bound the covertness metric, put a direct constraint on $\V{\Qinfn, \Qinnn}$, and show the remaining term vanishes exponentially fast. First note that, by the triangle inequality, we have
\begin{align}
\label{eq:V-ch-rsl-tri-ineq-1}
    \V{\Qcoden,\Qinnn}\leq\V{\Qcoden, \Qinfn}+\V{\Qinfn, \Qinnn}.
\end{align}
We first analyze the second term on the left-hand side of \eqref{eq:V-ch-rsl-tri-ineq-1}. By the basic property of the variational distance, we have
\begin{align}
  &\V{\Qinfn, \Qinnn}= \P[\Qinfn]{\Qinfn\left(\vct{Z}^n\right)\geq\Qinnn\left(\vct{Z}^n\right)} - \P[\Qinnn]{\Qinfn\left(\vct{Z}^n\right)\geq\Qinnn\left(\vct{Z}^n\right)}\\
    \label{eq:V-ch-rsl-cvt-sp-1}
  &\quad=\P[\Qinfn]{\sum_{i=1}^n\log\frac{\Qinf\left(\vct{Z}_i\right)}{\Qinn\left(\vct{Z}_i\right)}\geq0} - \P[\Qinnn]{\sum_{i=1}^n\log\frac{\Qinf\left(\vct{Z}_i\right)}{\Qinn\left(\vct{Z}_i\right)}\geq0}\\
    \label{eq:V-ch-rsl-cvt-sp-2}
  &\quad\labrel={eq:V-ch-rsl-cvt-subch-split}\P[\parQinfn]{\sum_{i=1}^n\sum_{j=1}^m\log\frac{\parjQinf\left(\tilde{Z}_{ij}\right)}{\parjQinn\left(\tilde{Z}_{ij}\right)}\geq0}- \P[\parQinnn]{\sum_{i=1}^n\sum_{j=1}^m\log\frac{\parjQinf\left(\tilde{Z}_{ij}\right)}{\parjQinn\left(\tilde{Z}_{ij}\right)}\geq0},
\end{align}
where \eqref{eq:V-ch-rsl-cvt-subch-split} follows since we apply the orthogonal transform $\vct{U}_w^{\prime\hermittian}$ to the per-channel-use observation $\vct{Z}_i$, and this mapping does not reduce the variational distance (i.e., the equality of data-processing inequality holds). Note that after the orthogonal transform, we could truncate the last $N_w-m$ components, since they contain pure noise (as they correspond to the null space of $\vct{H}_w^\hermittian$).
Then, for every $i\in\intseq{1}{n}$,
\begin{align}
  \mu_{1j}&\eqdef\expect[\parjQinf]{\log\frac{\parjQinf\left(Z_i\right)}{\parjQinn\left(Z_i\right)}}\\
            &=\expect[\parjQinf]{-\frac{\lambda_{w,j}^2\rho_{n,j}}{2\sigma_w^2}+\blog{\cosh\left(\frac{\lambda_{w,j}a_{n,j}Z_i}{\sigma_w^2}\right)}}\\
    &=\frac{\lambda_{w,j}^4\rho_{n,j}^2}{4\sigma_w^4}+\calO\left(\rho_{n,j}^3\right),
\end{align}
\begin{align}
  \sigma_{1j}^2&\eqdef\Var{\log\frac{\parjQinf\left(Z_i\right)}{\parjQinn\left(Z_i\right)}}\\
  &=\expect[\parjQinf]{\log^2\frac{\parjQinf\left(Z_i\right)}{\parjQinn\left(Z_i\right)}}-\calO\left(\rho_{n,j}^4\right)\\
    &=\frac{\lambda_{w,j}^4\rho_{n,j}^2}{2\sigma_w^4}+\calO\left(\rho_{n,j}^3\right),
\end{align}
\begin{align}
    t_{1j}\eqdef\expect[\parjQinf]{\left|\log\frac{\parjQinf\left(Z_i\right)}{\parjQinn\left(Z_i\right)}-\mu_{1j}\right|^3}=\calO\left(\rho_{n,j}^3\right).
\end{align}
Therefore, by the Berry-Esseen Theorem, we have 
\begin{align}
    &\quad\P[\parQinfn]{\sum_{i=1}^n\sum_{j=1}^m\log\frac{\parjQinf\left(\tilde{Z}_{ij}\right)}{\parjQinn\left(\tilde{Z}_{ij}\right)}\geq0}
    \leq\Qfunc{-\sqrt{\frac{n}{2}\sum_{j=1}^m\frac{\lambda_{w,j}^4\rho_{n,j}^2}{4\sigma_w^4}}}+\frac{6n\sum_{j=1}^mt_{1j}}{\left(n\sum_{j=1}^m\sigma_{1j}^2\right)^{3/2}}\\
    &=1-\Qfunc{\sqrt{\frac{n}{2}\sum_{j=1}^m\frac{\lambda_{w,j}^4\rho_{n,j}^2}{4\sigma_w^4}}}+\calO\left(\frac{1}{\sqrt{n}}\right).
\end{align}
Similarly, for every $i\in\intseq{1}{n}$,
\begin{align}
    \mu_{0j}&\eqdef\expect[\parjQinn]{\log\frac{\parjQinf\left(Z_i\right)}{\parjQinn\left(Z_i\right)}}=-\frac{\lambda_{w,j}^4\rho_{n,j}^2}{4\sigma_w^4}+\calO\left(\rho_{n,j}^3\right),\\
  \sigma_{0j}^2&\eqdef\Var{\log\frac{\parjQinf\left(Z_i\right)}{\parjQinn\left(Z_i\right)}}
                 =\frac{\lambda_{w,j}^4\rho_{n,j}^2}{2\sigma_w^4}+\calO\left(\rho_{n,j}^3\right),\\
    t_{0j}&\eqdef\expect[\parjQinn]{\left|\log\frac{\parjQinf\left(Z_i\right)}{\parjQinn\left(Z_i\right)}-\mu_{1j}\right|^3}=\calO\left(\rho_{n,j}^3\right).
\end{align}
We therefore have
\begin{align}
    &\quad\P[\parQinnn]{\sum_{i=1}^n\sum_{j=1}^m\log\frac{\parjQinf\left(\tilde{Z}_{ij}\right)}{\parjQinn\left(\tilde{Z}_{ij}\right)}\geq0}\nonumber\\
    &\geq\Qfunc{\sqrt{\frac{n}{2}\sum_{j=1}^m\frac{\lambda_{w,j}^4\rho_{n,j}^2}{4\sigma_w^4}}}-\frac{6n\sum_{j=1}^mt_{0j}}{\left(n\sum_{j=1}^m\sigma_{0j}^2\right)^{3/2}}\\
    &=\Qfunc{\sqrt{\frac{n}{2}\sum_{j=1}^m\frac{\lambda_{w,j}^4\rho_{n,j}^2}{4\sigma_w^4}}}-\calO\left(\frac{1}{\sqrt{n}}\right).
\end{align}
Eventually, we find an upper bound for \eqref{eq:V-ch-rsl-cvt-sp-2} as follows:
\begin{align}
\label{eq:V-ch-rsl-cvt-delta}
  \V{\Qinfn,\Qinnn}\leq 1-2\Qfunc{\sqrt{\frac{n}{2}\sum_{j=1}^m\frac{\lambda_{w,j}^4\rho_{n,j}^2}{4\sigma_w^4}}}
  +\calO\left(\frac{1}{\sqrt{n}}\right)\leq \delta-\frac{1}{\sqrt{n}},
\end{align}
where the $-\frac{1}{\sqrt{n}}$ term is added to ensure that the $\V{\Qinfn, \Qinnn}$ term is less than $\delta$ for $n$ large enough.
Equivalently, we impose a covertness constraint
\begin{align}
    \label{eq:V-ch-rsl-cvt-cnst-1}
    \frac{1}{4\sigma_w^4}\tr{\vct{\Lambda}_w\vct{T}\vct{\Lambda}_w^\hermittian\vct{\Lambda}_w\vct{T}\vct{\Lambda}_w^\hermittian}\leq 2 - \frac{C}{\sqrt{n}\invQdelta}, 
\end{align}
for some $C>0$ and the constraint \eqref{eq:V-ch-rsl-cvt-cnst-1} would be the main concern in the power design optimization.
Combining \eqref{eq:V-ch-rsl-tri-ineq-1} with \eqref{eq:V-ch-rsl-cvt-delta}, we therefore have, for $n$ large enough,
\begin{align}
\label{eq:V-ch-rsl-cvt-close-delta}
  \V{\Qcoden,\Qinnn}&\leq\V{\Qcoden, \Qinfn}
  + \V{\Qinfn, \Qinnn}\nonumber\\
  &\leq \V{\Qcoden,\Qinfn}+\delta-\frac{1}{\sqrt{n}}.
\end{align}
For the term $\V{\Qcoden, \Qinfn}$ in \eqref{eq:V-ch-rsl-cvt-close-delta}, our analysis follows from \cite[Lemma 5]{Bloch2016}, and we recall the following lemma. 
\begin{lemma}
For any $\theta>0$, 
\begin{align}
\label{eq:V-ch-rsl-code-dist}
  \expect[]{\V{\Qcoden,\Qinfn}}\leq&\P[\PiQnn\Wwn]{\log\frac{\Wwn\left(\vct{Z}^n|\vct{X}^n\right)}{\Qinnn\left(\vct{Z}^n\right)}\geq\theta}+ \frac{1}{2}\sqrt{\frac{e^\theta}{M_nK_n}}.
\end{align}
\end{lemma}
Similarly, we apply the orthogonal transform $\vct{U}_w^{\prime\hermittian}$ to the observations and decompose them into observations on each sub-channel; we obtain
\begin{equation}
    \log\frac{\Wwn\left(\vct{Z}^n|\vct{X}^n\right)}{\Qinnn\left(\vct{Z}^n\right)}=\sum_{i=1}^n\sum_{j=1}^m\log\frac{\parjWw\left(\tilde{Z}_{ij}|\tilde{X}_{ij}\right)}{\parjQinn\left(\tilde{Z}_{ij}\right)}.
\end{equation}
Since $\tilde{x}_{ij}\in\{-a_{n,j},a_{n,j}\}$,
$\sum_{j=1}^m\log\frac{\parjWw\left(\tilde{Z}_{ij}|\tilde{X}_{ij}=\tilde{x}_{ij}\right)}{\parjQinn\left(\tilde{Z}_{ij}\right)}$ $\sim\sum_{j=1}^m\N{\frac{\lambda_{w,j}^2\rho_{n,j}}{2\sigma_w^2}}{\frac{\lambda_{w,j}^2\rho_{n,j}}{\sigma_w^2}}.$
Therefore, by setting 
    $\theta=\left(1+\epsilon\right)n\sum_{j=1}^m\frac{\lambda_{w,j}^2\rho_{n,j}}{2\sigma_w^2},$
and using Hoeffding's inequality, we have $\P[\PiPnn\Wwn]{\log\frac{\Wwn\left(\vct{Z}^n|\vct{X}^n\right)}{\Qinnn\left(\vct{Z}^n\right)}\geq\theta}$$\leq\exp\left(-\frac{\epsilon^2n\sum_{j=1}^m\lambda_{w,j}^2\rho_{n,j}}{8\sigma_w^2}\right).$
Therefore, 
    $\expect[]{\V{\Qcoden,\Qinfn}}\leq\exp\left(-\frac{\epsilon^2n\sum_{j=1}^m\lambda_{w,j}^2\rho_{n,j}}{8\sigma_w^2}\right) + \frac{1}{2}\sqrt{\frac{e^\theta}{M_nK_n}}.$
Eventually, recalling \eqref{eq:achv-V-rho-def}, if we choose
\begin{align}
    \label{eq:V-ch-rsl-code-size}
    \log M_nK_n=\left(1+\omega\right)\left(1+\epsilon\right)n\sum_{j=1}^m\frac{\lambda_{w,j}^2\rho_{n,j}}{2\sigma_w^2}\nonumber\\
    =\left(1+\xi\right)\sum_{j=1}^m\frac{\lambda_{w,j}^2\tau_j\sqrt{n}\invQdelta}{2\sigma_w^2},
\end{align}
then 
\begin{equation}
\label{eq:V-ch-rsl-code-dist-exp}
    \expect[]{\V{\Qcoden,\Qinfn}}\leq e^{-\theta_2\sqrt{n}\invQdelta},
\end{equation}
for some appropriate choice of $\theta_2>0$. The result follows by combining \eqref{eq:V-ch-rsl-cvt-close-delta} and \eqref{eq:V-ch-rsl-code-dist-exp}. 
\end{IEEEproof}

\section{Proof of \Cref{lemma:V-compound-approximation}}
\label{sec:proof-lm-V-compound-approximation}

\begin{IEEEproof}
  We start by using $\V{\Qcoden_{\vct{H}}, \Qinnn}$ to approximate $\V{\Qcoden_{\widetilde{\vct{H}}}, \Qinnn}$. 
  For a fixed $\vct{P}_n$, by the triangle inequality,
\begin{align}
\label{eq:V-cmp-decompose}
    \abs{\V{\Qcoden_{\widetilde{\vct{H}}}, \Qinnn}-\V{\Qcoden_{\vct{H}}, \Qinnn}}
    \leq\V{\Qcoden_{\widetilde{\vct{H}}}, \Qcoden_{\vct{H}}}.
\end{align}
Note that for a given code $\calC$ and $\vct{x}^{(\ell k)n}\in\calC$,
\begin{align}
  \V{\Qcoden_{\widetilde{\vct{H}}}, \Qcoden_{\vct{H}}}&\leq\frac{\sum_{\ell=1}^{M_n}\sum_{k=1}^{K_n}}{M_nK_n}\V{\tWwn\left(\smash{\vct{Z}^n|\vct{x}^{(\ell k)n}}\right),\Wwn\left(\smash{\vct{Z}^n|\vct{x}^{(\ell k)n}}\right)}\\
    \label{eq:V-cmp-code-dist-mkcode}
    &\labrel={eq:V-cmp-code-dist-par}\frac{\sum_{\ell=1}^{M_n}\sum_{k=1}^{K_n}}{M_nK_n}\V{\partWwn\left(\smash{\parVarZ^n|\parVarx^{(\ell k)n}}\right),\parWwn\left(\smash{\parVarZ^n|\parVarx^{(\ell k)n}}\right)},
\end{align}
where \eqref{eq:V-cmp-code-dist-par} follows since the orthogonal transformation preserves the variational distance. To characterize the behavior of $\V{\partWwn\left(\smash{\parVarZ^n|\parVarx^{(\ell k)n}}\right),\parWwn\left(\smash{\parVarZ^n|\parVarx^{(\ell k)n}}\right)}$, we follow the proof of \cite[Lemma 5]{He2014} and consider a specific codeword $\parVarx^n$, 
\begin{align}
    &\quad2\V{\partWwn\left(\smash{\parVarZ^n|\parVarx^n}\right),\parWwn\left(\smash{\parVarZ^n|\parVarx^n}\right)}=\int_{\parVarz^n}\left|\partWwn\left(\smash{\parVarz^n|\parVarx^n}\right)-\parWwn\left(\smash{\parVarz^n|\parVarx^n}\right)\right|d\parVarz^n\\
    &=\int_{\sum_{i=1}^n\norm[2]{\parVarz_i-\widetilde{\vct{\Lambda}}\parVarx_i}^2\geq r_n^2}\left|\partWwn\left(\smash{\parVarz^n|\parVarx^n}\right)-\parWwn\left(\smash{\parVarz^n|\parVarx^n}\right)\right|d\parVarz^n\nonumber\\
    &\quad+\int_{\sum_{i=1}^n\norm[2]{\parVarz_i-\widetilde{\vct{\Lambda}}\parVarx_i}^2< r_n^2}\left|\partWwn\left(\smash{\parVarz^n|\parVarx^n}\right)-\parWwn\left(\smash{\parVarz^n|\parVarx^n}\right)\right|d\parVarz^n\\
    \label{eq:V-cmp-code-dist-decom-1}
    &\leq \P[\partWwn]{\sum_{i=1}^n\norm[2]{\parVarz_i-\widetilde{\vct{\Lambda}}\parVarx_i}^2\geq r_n^2} + \P[\parWwn]{\sum_{i=1}^n\norm[2]{\parVarz_i-\widetilde{\vct{\Lambda}}\parVarx_i}^2\geq r_n^2}\nonumber\\ 
    &\quad+\int_{\sum_{i=1}^n\norm[2]{\parVarz_i-\widetilde{\vct{\Lambda}}\parVarx_i}^2< r_n^2}\left|\partWwn\left(\smash{\parVarz^n|\parVarx^n}\right)-\parWwn\left(\smash{\parVarz^n|\parVarx^n}\right)\right|d\parVarz^n,
\end{align}
where $r_n\eqdef \sqrt{nm(1+\epsilon)\sigma_w^2}+\epsilon_n\norm[F]{\parVarx^n}$, which is close to $\sqrt{nm(1+\epsilon)\sigma_w^2}$ as $n$ grows to infinity, and $\epsilon\in(0,1)$.
%
Note that the first term of \eqref{eq:V-cmp-code-dist-decom-1} can be bounded by the concentration inequality for the sub-exponential random variables as follows:
\begin{align}
  &\P[\partWwn]{\sum_{i=1}^n\norm[2]{\parVarz_i-\widetilde{\vct{\Lambda}}\parVarx_i}^2\geq r_n^2}\leq\exp\left(-\frac{(r_n^2-nm\sigma_w^2)^2}{8nm}\right)=\exp\left(-\calO\left(n\right)\right).
\end{align}
Also, by the triangle inequality for the Frobenius norm, we have
\begin{align}
    \norm[F]{\parVarz^n-\widetilde{\vct{\Lambda}}\parVarx^n}&\leq\norm[F]{\strut\parVarz^n-\vct{\Lambda}\parVarx^n} + \norm[F]{\left(\vct{\Lambda}-\widetilde{\vct{\Lambda}}\right)\parVarx^n}\\
    &\leq\norm[F]{\strut\parVarz^n-\vct{\Lambda}\parVarx^n} + \sqrt{\epsilon_n^2\sum_{i=1}^n\norm[2]{\parVarx_i}^2}\\
    &=\norm[F]{\strut\parVarz^n-\vct{\Lambda}\parVarx^n} + \epsilon_n\norm[F]{\parVarx^n}.
\end{align}
Therefore, $\sum_{i=1}^n\norm[2]{\parVarz_i-\widetilde{\vct{\Lambda}}\parVarx_i}^2\geq r_n^2$ implies that
  $\sum_{i=1}^n\norm[2]{\strut\parVarz_i-\vct{\Lambda}\parVarx_i}^2\geq \left(r_n-\epsilon_n\norm[F]{\parVarx^n}\right)^2=nm(1+\epsilon)\sigma_w^2.$
Accordingly, we can use the concentration inequality for the sub-exponential random variables and obtain
$\P[\parWwn]{\sum_{i=1}^n\norm[2]{\parVarz_i-\widetilde{\vct{\Lambda}}\parVarx_i}^2\geq r_n^2}\leq\exp\left(-\frac{nm\epsilon^2}{8}\right).$
We next investigate the variation between densities $\parWwn$ and $\partWwn$ caused by the difference between $\vct{H}$ and $\widetilde{\vct{H}}$ as follows:
\begin{align}
    \left|\log\frac{\parWwn\left(\smash{\parVarz^n|\parVarx^n}\right)}{\partWwn\left(\smash{\parVarz^n|\parVarx^n}\right)}\right|&=\frac{1}{2\sigma_w^2}\left|\sum_{i=1}^n\left(\norm[2]{\smash{\parVarz_i-\widetilde{\vct{\Lambda}}\parVarx_i}}^2-\norm[2]{\parVarz_i-\vct{\Lambda}\parVarx_i}^2\right)\right|\\
    &\labrel\leq{eq:V-cmp-code-dist-dset}\frac{1}{2\sigma_w^2}\left(\left|\sum_{i=1}^n\sum_{j=1}^m2(\tilde{z}_{ij}-\widetilde{\lambda}_j\tilde{x}_{ij})(\epsilon_n\tilde{x}_{ij})\right|+\left|\sum_{i=1}^n\sum_{j=1}^m(\epsilon_n\tilde{x}_{ij})^2\right|\right)\\
    \label{eq:V-cmp-code-dist-ub-1}
    &\labrel\leq{eq:V-cmp-code-dist-cauchy} \frac{1}{\sigma_w^2}\sqrt{\sum_{i=1}^n \norm[2]{\parVarz_{i}-\widetilde{\vct{\Lambda}}\parVarx_{i}}^2 \epsilon_n^2\norm[F]{\parVarx^n}^2}+\frac{1}{2\sigma_w^2}\epsilon_n^2\norm[F]{\parVarx^n}^2,
\end{align}
where \eqref{eq:V-cmp-code-dist-dset} follows from the definition of $\calS_{J,n}$ and the triangle inequality, and \eqref{eq:V-cmp-code-dist-cauchy} follows from the Cauchy-Schwartz inequality.
Then, for the last term of \eqref{eq:V-cmp-code-dist-decom-1}, we proceed as follows:
\begin{align}
    &\quad\int_{\sum_{i=1}^n\norm[2]{\parVarz_i-\widetilde{\vct{\Lambda}}\parVarx_i}^2< r_n^2}\left|\partWwn\left(\smash{\parVarz^n|\parVarx^n}\right)-\parWwn\left(\smash{\parVarz^n|\parVarx^n}\right)\right|d\parVarz^n \nonumber\\
    &\labrel\leq{eq:V-cmp-code-dist-ub-2}\int_{\sum_{i=1}^n\norm[2]{\parVarz_i-\widetilde{\vct{\Lambda}}\parVarx_i}^2< r_n^2}\partWwn\left(\smash{\parVarz^n|\parVarx^n}\right)\left(f_n + \calO\left(f_n^2\right)\right)d\parVarz^n\\
    &\labrel\leq{eq:V-cmp-code-dist-ub-3} f_n+\calO\left(f^2_n\right) = \calO\left(n^{\frac{1}{2}}\norm[F]{\parVarx^n}e^{-n\log 2}\right),
\end{align}
where we let $f_n=\frac{1}{\sigma_w^2}\sqrt{r_n^2 \epsilon_n^2\norm[F]{\parVarx^n}^2}+\frac{1}{2\sigma_w^2}\epsilon_n^2\norm[F]{\parVarx^n}^2$, since $\sum_{i=1}^n\norm[2]{\parVarz_i-\widetilde{\vct{\Lambda}}\parVarx_i}^2< r_n^2$, \eqref{eq:V-cmp-code-dist-ub-2} and \eqref{eq:V-cmp-code-dist-ub-3} follow since 
    $\left|1-\frac{\parWwn\left(\smash{\parVarz^n|\parVarx^n}\right)}{\partWwn\left(\smash{\parVarz^n|\parVarx^n}\right)}\right|\leq\max\{{e^{f_n}-1,1-e^{-f_n}}\}\leq f_n + \calO\left(f_n^2\right).$
Therefore, we have, for $n$ large enough,
    $\V{\partWwn\left(\smash{\parVarZ^n|\parVarx^n}\right),\parWwn\left(\smash{\parVarZ^n|\parVarx^n}\right)}\leq\calO\left(n^{\frac{1}{2}}\norm[F]{\parVarx^n}e^{-n\log 2}\right),$
and from \eqref{eq:V-cmp-code-dist-mkcode}, we obtain 
\begin{align}
    \V{\Qcoden_{\widetilde{\vct{H}}},\Qcoden_{\vct{H}}}\leq\sum_{\ell=1}^{M_n}\sum_{k=1}^{K_n}\frac{1}{M_nK_n}\calO\left(n^{\frac{1}{2}}\norm[F]{\parVarx^{(\ell k)n}}e^{-n\log 2}\right).
\end{align}
For any \ac{BPSK} code generated independently according to $\PiPnn$, the power of generated codewords are fixed, and therefore we have 
\begin{align}
\label{eq:V-cmp-code-dist-ub-4}
  \V{\Qcoden_{\widetilde{\vct{H}}},\Qcoden_{\vct{H}}}\leq\calO\left(n^{\frac{3}{4}}e^{-n\log 2}\right).
\end{align}

Eventually, combining \eqref{eq:V-cmp-decompose} with \eqref{eq:V-cmp-code-dist-ub-4}, the result follows.
\end{IEEEproof}

%





\ifCLASSOPTIONcaptionsoff
  \newpage
\fi



\bibliographystyle{IEEEtran}
\bibliography{Bibliography/IT,Bibliography/Math,Bibliography/Communication}
\end{document}

%% file: arcom_functions.tex
\renewcommand{\P}[2][]{{\mathbb{P}_{#1}}{\left(#2\right)}} 
\newcommand{\Var}[1]{{\text{\textnormal{Var}}{\left(#1\right)}}} 
\newcommand{\avgD}[3][]{{\mathbb{D}_{#1}}\!\left(\smash{#2}\,\middle\Vert\,\smash{#3}\right)} 
\newcommand{\V}[1]{{{\mathbb{V}}\!\left(\smash{#1}\right)}} 
\newcommand{\avgI}[1]{{{\mathbb{I}}\!\left(#1\right)}} 
\newcommand{\avgH}[1]{{\mathbb{H}}\!\left(#1\right)}

\newcommand{\Hb}[1]{{h_b}\left(#1\right)} 
\newcommand{\h}[1]{{{h}}\left(#1\right)} 

\newcommand{\card}[1]{\ensuremath{\left|{#1}\right|}} 
\newcommand{\abs}[1]{\ensuremath{\left|#1\right|}} 
\newcommand{\norm}[2][]{\ensuremath{{\left\Vert{#2}\right\Vert}_{#1}}} 
\newcommand{\eqdef}{\ensuremath{\triangleq}} 
\newcommand{\intseq}[2]{\ensuremath{\llbracket{#1},{#2}\rrbracket}} 
\newcommand{\indic}[1]{\ensuremath{\mathds{1}\!\left\{#1\right\}}} 
\renewcommand{\leq}{\leqslant} 
\renewcommand{\geq}{\geqslant} 

\newcommand{\tr}[1]{\ensuremath{\text{\textnormal{tr}}\left(#1\right)}}  
\newcommand{\rk}[1]{\ensuremath{\text{\textnormal{rank}}\left(#1\right)}}  
\renewcommand{\det}[1]{{\left|{#1}\right|}}                              

\newcommand{\expect}[2][]{\ensuremath{\mathbb{E}_{#1}{\left\{#2\right\}}}}

\newcommand{\vct}[1]{\ensuremath{\mathbf{#1}}} 

\newcommand{\limn}[2][n\rightarrow\infty]{\underset{#1}{\mbox{\textnormal{lim}}\;}{#2}}
\renewcommand{\liminf}[2][n\rightarrow\infty]{\underset{#1}{\mbox{\textnormal{liminf}}\;}{#2}}
\renewcommand{\limsup}[2][n\rightarrow\infty]{\underset{#1}{\mbox{\textnormal{limsup}}\;}{#2}}
\newcommand{\N}[2]{\calN\left({#1},{#2}\right)}
\newcommand{\blog}[1]{\mbox{\textnormal{log}}\left({#1}\right)}
\newcommand{\Qfunc}[1]{Q\left({#1}\right)}
\newcommand{\invQfunc}[1]{Q^{-1}\left({#1}\right)}
\newcommand{\matvct}[3]{\ensuremath{\left[#1_{#2}~\dots~#1_{#3}\right]}}
\newcommand{\diag}[1]{\textnormal{diag}\left(#1\right)}
\newcommand{\floor}[1]{\lfloor#1\rfloor}
\newcommand{\ceil}[1]{\left\lceil#1\right\rceil}

%% file: arcom_notation.tex
\DeclareMathAlphabet{\eurm}{U}{eur}{m}{n}
\DeclareMathAlphabet{\mathbsf}{OT1}{cmss}{bx}{n}
\DeclareMathAlphabet{\mathssf}{OT1}{cmss}{m}{sl}
\DeclareMathAlphabet{\mathcsf}{OT1}{cmss}{sbc}{n}

\newcommand{\randomvalue}[1]{{\uppercase{#1}}}

\DeclareSymbolFont{bsfletters}{OT1}{cmss}{bx}{n}  
\DeclareSymbolFont{ssfletters}{OT1}{cmss}{m}{n}
\DeclareMathSymbol{\bsfGamma}{0}{bsfletters}{'000}
\DeclareMathSymbol{\ssfGamma}{0}{ssfletters}{'000}
\DeclareMathSymbol{\bsfDelta}{0}{bsfletters}{'001}
\DeclareMathSymbol{\ssfDelta}{0}{ssfletters}{'001}
\DeclareMathSymbol{\bsfTheta}{0}{bsfletters}{'002}
\DeclareMathSymbol{\ssfTheta}{0}{ssfletters}{'002}
\DeclareMathSymbol{\bsfLambda}{0}{bsfletters}{'003}
\DeclareMathSymbol{\ssfLambda}{0}{ssfletters}{'003}
\DeclareMathSymbol{\bsfXi}{0}{bsfletters}{'004}
\DeclareMathSymbol{\ssfXi}{0}{ssfletters}{'004}
\DeclareMathSymbol{\bsfPi}{0}{bsfletters}{'005}
\DeclareMathSymbol{\ssfPi}{0}{ssfletters}{'005}
\DeclareMathSymbol{\bsfSigma}{0}{bsfletters}{'006}
\DeclareMathSymbol{\ssfSigma}{0}{ssfletters}{'006}
\DeclareMathSymbol{\bsfUpsilon}{0}{bsfletters}{'007}
\DeclareMathSymbol{\ssfUpsilon}{0}{ssfletters}{'007}
\DeclareMathSymbol{\bsfPhi}{0}{bsfletters}{'010}
\DeclareMathSymbol{\ssfPhi}{0}{ssfletters}{'010}
\DeclareMathSymbol{\bsfPsi}{0}{bsfletters}{'011}
\DeclareMathSymbol{\ssfPsi}{0}{ssfletters}{'011}
\DeclareMathSymbol{\bsfOmega}{0}{bsfletters}{'012}
\DeclareMathSymbol{\ssfOmega}{0}{ssfletters}{'012}

\newcommand{\rvS}{{\randomvalue{S}}}	
\newcommand{\rvW}{{\randomvalue{W}}}	

\newcommand{\calC}{{\mathcal{C}}}

\newcommand{\calE}{{\mathcal{E}}}

\newcommand{\calM}{{\mathcal{M}}}
\newcommand{\calN}{{\mathcal{N}}}
\newcommand{\calO}{{\mathcal{O}}}

\newcommand{\calS}{{\mathcal{S}}}

\newcommand{\calX}{{\mathcal{X}}}
\newcommand{\calY}{{\mathcal{Y}}}

\newcommand{\calZ}{{\mathcal{Z}}}

\newcommand{\bbN}{\mathbb{N}}
\newcommand{\bbP}{\mathbb{P}}
\newcommand{\bbR}{\mathbb{R}}

\newcommand{\Pinn}{P_0}
\newcommand{\Pinf}{P_n}
\newcommand{\Qinn}{Q_0}
\newcommand{\Qinf}{Q_n}
\newcommand{\Qcode}{\widehat{Q}}

\newcommand{\QinfH}{Q_{\vct{H}}}
\newcommand{\QinfHn}{\QinfH^{\otimes n}}

\newcommand{\QinftH}{Q_{\widetilde{\vct{H}}}}
\newcommand{\QinftHn}{\QinftH^{\otimes n}}
\newcommand{\QinfHw}{Q_{\vct{H}_w}}
\newcommand{\QinfHwn}{\QinfHw^{\otimes n}}

\newcommand{\PiQn}{\Pi_{\vct{Q}_n}}
\newcommand{\PiPn}{\Pi_{\vct{P}_n}}
\newcommand{\PiQnn}{\Pi_{\vct{Q}_n}^{\otimes n}}
\newcommand{\PiPnn}{\Pi_{\vct{P}_n}^{\otimes n}}
\newcommand{\Pinnn}{P_0^{\otimes n}}
\newcommand{\Pinfn}{P_n^{\otimes n}}
\newcommand{\Qinnn}{Q_0^{\otimes n}}
\newcommand{\Qinfn}{Q_n^{\otimes n}}
\newcommand{\Qcoden}{\widehat{Q}^n}

\newcommand{\parPinn}{\widetilde{P}_0}
\newcommand{\parPinf}{\widetilde{P}_n}
\newcommand{\parQinn}{\widetilde{Q}_0}
\newcommand{\parQinf}{\widetilde{Q}_n}

\newcommand{\parPinfn}{\widetilde{P}_n^{\otimes n}}
\newcommand{\parQinnn}{\widetilde{Q}_0^{\otimes n}}
\newcommand{\parQinfn}{\widetilde{Q}_n^{\otimes n}}

\newcommand{\parjPinn}{\widetilde{P}_{0,j}}
\newcommand{\parjPinf}{\widetilde{P}_{n,j}}
\newcommand{\parjQinn}{\widetilde{Q}_{0,j}}
\newcommand{\parjQinf}{\widetilde{Q}_{n,j}}

\newcommand{\parjPinnn}{\widetilde{P}_{0,j}^{\otimes n}}

\newcommand{\parVarX}{\tilde{\vct{X}}}
\newcommand{\parVarY}{\tilde{\vct{Y}}}
\newcommand{\parVarZ}{\tilde{\vct{Z}}}
\newcommand{\parVarx}{\tilde{\vct{x}}}

\newcommand{\parVarz}{\tilde{\vct{z}}}

\newcommand{\hermittian}{\intercal}

\newcommand{\Ww}{W_{\vct{Z}|\vct{X}}}
\newcommand{\Wwn}{W_{\vct{Z}|\vct{X}}^{\otimes n}}
\newcommand{\tWwn}{\widetilde{W}_{\vct{Z}|\vct{X}}^{\otimes n}}

\newcommand{\Wb}{W_{\vct{Y}|\vct{X}}}
\newcommand{\Wbn}{W_{\vct{Y}|\vct{X}}^{\otimes n}}
\newcommand{\parWw}{W_{\tilde{\vct{Z}}|\tilde{\vct{X}}}}
\newcommand{\parWwn}{W_{\tilde{\vct{Z}}|\tilde{\vct{X}}}^{\otimes n}}

\newcommand{\partWwn}{\widetilde{W}_{\tilde{\vct{Z}}|\tilde{\vct{X}}}^{\otimes n}}
\newcommand{\parWb}{W_{\tilde{\vct{Y}}|\tilde{\vct{X}}}}

\newcommand{\parjWw}{W_{\tilde{Z}^{(j)}|\tilde{X}^{(j)}}}

\newcommand{\parjWb}{W_{\tilde{Y}^{(j)}|\tilde{X}^{(j)}}}
\newcommand{\parjWbn}{W_{\tilde{Y}^{(j)}|\tilde{X}^{(j)}}^{\otimes n}}

\newcommand{\invQdelta}{\invQfunc{\frac{1-\delta}{2}}}

\newcommand{\Null}{\textnormal{Null}}

%% file: arcom_acronyms.tex
\acrodef{AEP}{Asymptotic Equipartition Property}
\acrodef{AoA}{Angle of Arrival}
\acrodef{AWGN}{Additive White Gaussian Noise}
\acrodef{AVC}[AVC]{Arbitrarily Varying Channel}
\acrodef{BER}{Bit-Error-Rate}
\acrodef{BEC}{Binary Erasure Channel}
\acrodef{BPSK}{Binary Phase-Shift Keying}
\acrodef{BSC}{Binary Symmetric Channel}
\acrodef{BICM}[BICM]{Bit-Interleaved Coded-Modulation}
\acrodef{BIDMC}[BI-DMC]{Binary-Input Discrete Memoryless Channel}
\acrodef{CDF}[CDF]{Cumulative Distribution Function}
\acrodef{CGF}[CGF]{Cumulant Generating Function}
\acrodef{CLT}[CLT]{Central Limit Theorem}
\acrodef{CSI}[CSI]{Channel State Information}
\acrodef{DMC}[DMC]{Discrete Memoryless Channel}
\acrodef{DMS}[DMS]{Discrete Memoryless Source}
\acrodef{FER}[FER]{Frame Error Rate}
\acrodef{iid}[i.i.d.]{independent and identically distributed}
\acrodef{IoT}[IoT]{Internet of Things}
\acrodef{LPD}[LPD]{Low Probability of Detection}
\acrodef{LDPC}[LDPC]{Low-Density Parity-Check}
\acrodef{MAC}[MAC]{multiple-access channel}
\acrodef{MGF}[MGF]{Moment Generating Function}
\acrodef{MLC}[MLC]{Multi-Level Coding}
\acrodef{MIMO}[MIMO]{Multiple-Input Multiple-Output}
\acrodef{MISO}{Multiple-Input Single-Output}
\acrodef{MSD}[MSD]{Multi-Stage Decoding}
\acrodef{PDF}[PDF]{Probability Density Function}
\acrodef{PMF}[PMF]{Probability Mass Function}
\acrodef{PPM}[PPM]{Pulse-Position Modulation}
\acrodef{PSD}{Power Spectral Density}
\acrodef{PSK}{Phase Shift Keying}
\acrodef{QKD}{Quantum Key Distribution}
\acrodef{ROC}{Receiver Operating Characteristic}
\acrodef{CVQKD}{Continuous-Variable \ac{QKD}}
\acrodef{QPSK}{Quadrature Phase-Shift Keying}
\acrodef{RV}{random variable}
\acrodef{SIMO}{Single-Input Multiple-Output}
\acrodef{SISO}{Single-Input Single-Output}
\acrodef{SNR}{Signal-to-Noise Ratio}
\acrodef{TPCP}{Trace-Preserving Completely-Positive}
\acrodef{wrt}[w.r.t.]{with respect to}
\acrodef{WSS}{Wide Sense Stationary}
\acrodef{GSVD}{Generalized Singular Value Decomposition}
\acrodef{LLR}{Log-Likelihood Ratio}